\pdfoutput=1

\documentclass[11pt,twoside,a4paper,cmspaper,final,collab]{cms-tdr}
\def\svnVersion{280872}\def\cmsCernNo{2013-037}\def\cmsCernDate{\today}\def\cmsMessage{Published in Physical Review D as \href{http://dx.doi.org/10.1103/PhysRevD.91.052008}{\doi{10.1103/PhysRevD.91.052008.}}}

\begin{document}\cmsNoteHeader{SMP-12-017}

\hyphenation{had-ron-i-za-tion}
\hyphenation{cal-or-i-me-ter}
\hyphenation{de-vices}
\RCS$Revision: 280837 $
\RCS$HeadURL: svn+ssh://svn.cern.ch/reps/tdr2/papers/SMP-12-017/trunk/SMP-12-017.tex $
\RCS$Id: SMP-12-017.tex 280837 2015-03-16 15:01:33Z fabiocos $
\newlength\cmsFigWidth
\ifthenelse{\boolean{cms@external}}{\setlength\cmsFigWidth{0.4\textwidth}}{\setlength\cmsFigWidth{0.49\textwidth}}
\ifthenelse{\boolean{cms@external}}{\providecommand{\cmsLeft}{top}}{\providecommand{\cmsLeft}{left}}
\ifthenelse{\boolean{cms@external}}{\providecommand{\cmsRight}{bottom}}{\providecommand{\cmsRight}{right}}
\cmsNoteHeader{SMP-12-017}
\title{Measurements of jet multiplicity and differential production
cross sections of \texorpdfstring{\cPZ+jets events in proton-proton
collisions at $\sqrt{s}=7$\TeV}{Z+jets events in proton-proton
collisions at sqrt(s)=7 TeV}}

\date{\today}

\abstract{
Measurements of differential cross sections are presented for the
production of a \Z boson and at least one hadronic jet in
proton-proton collisions at $\sqrt{s} = 7$\TeV, recorded by the CMS
detector, using a data sample corresponding to an integrated
luminosity of 4.9\fbinv.  The jet multiplicity distribution is
measured for up to six jets. The differential cross sections are
measured as a function of jet transverse momentum and pseudorapidity for
the four highest transverse momentum jets. The distribution of the
scalar sum of jet transverse momenta is also measured as a function of
the jet multiplicity. The measurements are compared with theoretical
predictions at leading and next-to-leading order in perturbative QCD.
}

\hypersetup{%
pdfauthor={CMS Collaboration},%
pdftitle={Measurements of jet multiplicity and differential production cross sections of Z+jets events in proton-proton collisions at sqrt(s)=7 TeV},%
pdfsubject={CMS},%
pdfkeywords={CMS, physics, jets}}

\providecommand{\POWHEGBOX} {{\textsc{powheg-box}}\xspace}
\providecommand{\BLACKHAT} {{\textsc{BlackHat}}\xspace}
\providecommand{\RIVET} {{\textsc{rivet}}\xspace}

\maketitle
\section{Introduction}

Measurements of the production cross section of a \Z boson with one or
more jets in hadron collisions, hereafter \cPZ+jets, can be compared
with predictions of perturbative quantum chromodynamics (pQCD).
Analyses of data collected during the first run of the CERN LHC have
used two main theoretical approaches, developed in the last decade,
for the complete description of the associated production of
vector bosons and jets up to stable particles in the final
state. Multileg matrix elements, computed at leading order (LO) in
pQCD, have been combined with parton showers (PS), merging different
final jet multiplicities together. Alternatively, next-to-leading order
(NLO) matrix elements have been interfaced with parton showers for
final states of fixed jet multiplicity. CMS has relied on
\MADGRAPH~\cite{Alwall:2011uj} and
\POWHEGBOX~\cite{Nason:2004rx,Frixione:2007vw,Alioli:2010xd} as main
implementations of the former and latter approaches, respectively.  In
the last few years, novel techniques have been developed in order to merge
NLO calculations for several final state multiplicities in a
theoretically consistent way, and interface them with PS, as formerly
done for LO matrix elements. This approach may provide a NLO accuracy
for a range of complex topologies, overcoming the limitations of fixed
order NLO calculations, which cannot in general describe completely
inclusive distributions receiving contributions by final states of
different jet multiplicity. Furthermore, a description of these final
states up to stable particles is possible, since hadronization
models can be used in combination with these calculations. The
\cPZ+jets final state provides jet kinematic distributions that are
ideal for testing these different options for theoretical predictions.

Also, this process contributes a large background to many standard
model processes, like top production or diboson final states,
e.g. the associated production of a Higgs boson and a \Z, where the
former decays in \bbbar pairs and the second in charged
leptons~\cite{Chatrchyan:2013zna,Aad:2014xzb}.  Searches for phenomena
beyond the standard model may also be sensitive to this process, which
plays a particularly important role as the main background in the study of
supersymmetric scenarios with large missing transverse momentum. This
has been one of the main motivations for a previous analysis
of the angular distributions in \cPZ+jets events presented by
CMS~\cite{Chatrchyan:2013tna}, and the study presented in this paper
is complementing it with the measurement of jet spectra.

Measurements of \cPZ+jets production were published by the CDF and D0
collaborations based on a sample of proton-antiproton collisions at
$\sqrt{s}=1.96$\TeV~\cite{Aaltonen:2007ae,Abazov:2009av}, and by the
ATLAS~\cite{Aad:2011qv} and CMS~\cite{Chatrchyan:2011ne}
collaborations from a sample of proton-proton collisions at
$\sqrt{s}=7$\TeV collected at the LHC, corresponding to an integrated
luminosity of 0.036\fbinv. ATLAS has reported an updated measurement
at the same center-of-mass energy with a data set corresponding to an
integrated luminosity of 4.6\fbinv~\cite{Aad:2013ysa}.

In this paper, we update and expand upon the results obtained by the
CMS Collaboration at $\sqrt{s}=7$\TeV with a data sample corresponding
to an integrated luminosity of
$4.9\pm0.1\fbinv$~\cite{CMS-PAS-SMP-12-008} collected in 2011. We
present fiducial cross sections for \cPZ+jets production as a function
of the exclusive and inclusive jet multiplicity, where the \Z bosons
are identified through their decays into electron or muon pairs. The
contribution from \cPZ/$\gamma^{\ast}$ interference is considered to
be part of the measured signal. We measure the differential cross
sections as a function of the transverse momentum \pt and
pseudorapidity $\eta$ of the four highest-\pt jets in the event. The
pseudorapidity is defined as $\eta = - \ln \tan [\theta/2]$, where
$\theta$ is the polar angle with respect to the counterclockwise-rotating
proton beam. We also present results for the distribution of \HT, the
scalar sum of jet transverse momenta, measured as a function of the
inclusive jet multiplicity. The jet \pt and $\eta$ differential cross
sections are sensitive to higher order QCD corrections. \HT is an
observable characterizing globally the QCD emission structure of the
event, and it is often used as a discriminant variable in searches for
supersymmetric scenarios, to which \cPZ+jets contribute as a
background. The measurement of its distribution  is therefore of great
interest.

The paper is organized as follows. Section 2 presents a description of
the CMS apparatus and its main characteristics. Section 3 provides
details about the simulation used in this analysis. Section 4
discusses the event reconstruction and selection. Section 5 is devoted
to the estimation of the signal event selection efficiency and to the
subtraction of the background contributions. The procedure used to
correct the measurement for detector response and resolution is
presented in Section 6. Section 7 describes the estimation of the
systematic uncertainties, and in Section 8 the results are presented
and theoretical predictions are compared to them.

\section{The CMS detector}

The central feature of the CMS apparatus is a superconducting solenoid
of 6\unit{m} internal diameter that provides a
magnetic field of 3.8\unit{T}. The field volume
contains a silicon tracker, a lead tungstate crystal electromagnetic
calorimeter (ECAL), and a brass/scintillator hadron calorimeter; each
subdetector in the barrel section is enclosed by two end caps. The
magnet flux-return yoke is instrumented with gas-ionization tracking
devices for muon detection. In addition to the barrel and end cap
detectors, CMS has an extensive forward calorimetry system. CMS uses a
two-level trigger system. The first level is composed of custom
hardware processors, and uses local information from the calorimeters
and muon detectors to select the most interesting events in a fixed
time interval of less than 4\mus. The high-level trigger is a
processor farm that further decreases the event rate from a maximum of
100\unit{kHz} to roughly 300\unit{Hz}, before data storage. A detailed
description of the CMS detector can be found in
Ref.~\cite{Chatrchyan:2008zzk}.

Here we briefly outline the detector elements and performance
characteristics that are most relevant to this measurement.  The inner
tracker, which consists of silicon pixel and silicon strip detectors,
reconstructs charged-particle trajectories within the range
$\abs{\eta}< 2.5$.  The tracking system provides an impact parameter
resolution of 15\mum and a \pt resolution of 1.5\% for 100\GeV
particles. Energy deposits in the ECAL are matched to tracks in the
silicon detector and used to initiate the reconstruction algorithm for
electrons.  The tracking algorithm takes into account the energy lost
by electrons in the detector material through bremsstrahlung.  In the
energy range relevant for \Z-boson decays, the electron energy
resolution is below 3\%. Muon trajectories are reconstructed for
$\abs{\eta} < 2.4$ using detector planes based on three technologies:
drift tubes, cathode-strip chambers, and resistive-plate
chambers. Matching outer muon trajectories to tracks measured in the
silicon tracker provides an average \pt resolution of 1.6\% for the
\pt range used in this analysis. For the jets reconstructed in this
analysis, the \pt resolution is better than 10\% and the energy scale
uncertainty is less than 3\%~\cite{Chatrchyan:1369486}.

\section{Physics processes and detector simulation}
\label{data&mc}

Simulated events are used to correct the signal event yield for
detector effects and to subtract the contribution from background
events.  Simulated Drell--Yan \Z/$\gamma^{\ast}$, \ttbar, and \PW+jets
events are generated using the \MADGRAPH 5.1.1 \cite{Alwall:2011uj}
event generator. The package provides a tree level matrix-element
calculation with up to four additional partons in the final state for
vector boson production, and three additional partons for \ttbar
events. The leading-order CTEQ6L1 parton distribution functions
(PDF)~\cite{Pumplin:2002vw} are used with \MADGRAPH. The residual QCD
radiation, described by a parton shower algorithm, and the
hadronization, which turns the partons into a set of stable particles, are
implemented with \PYTHIA~6.424~\cite{Sjostrand:2006za} using the Z2
underlying event and fragmentation tune~\cite{Khachatryan:2010nk}. The default
$\alpha_{\mathrm{S}}$ value of the PDF set used is adopted for the event generator. 
The matrix-element and parton shower calculations are matched using the
\kt-MLM algorithm~\cite{Alwall:2007fs}.  Decays of the $\Pgt$ lepton
are described by the \TAUOLA 1.27 \cite{Golonka:2003xt}
package. Diboson events (\PW\PW, \PW\cPZ, \cPZ\cPZ) are modeled
entirely with \PYTHIA.  Single-top events in the $\PW\cPqt$ channel
are simulated using
\POWHEGBOX~\cite{Nason:2004rx,Frixione:2007vw,Alioli:2010xd,Re:2010bp},
and followed by \PYTHIA to describe QCD radiation beyond
NLO and hadronization.  An alternative
description of the Drell--Yan signal is used for the evaluation of
systematic uncertainties that is based on the
\SHERPA~1.4~\cite{Gleisberg:2008fv,Schumann:2007mg,Gleisberg:2008ta,Hoeche:2009rj}
tree level matrix-element calculation, which has up to four additional
partons in the final state, and uses the NLO
CTEQ6.6M~\cite{Nadolsky:2008zw} PDF set.

The total cross sections for the \Z signal and the $\PW$ background
are normalized to the next-to-next-to-leading-order (NNLO) predictions
that are obtained with \FEWZ~\cite{Melnikov:2006kv} and the
MSTW2008 \cite{Martin:2009iq} PDF set. The \ttbar cross section is
normalized to the NNLO prediction from
Ref.~\cite{Czakon:2013goa}. Diboson cross sections are rescaled to the
NLO predictions obtained with \MCFM~\cite{Campbell:2011bn}.

The interaction of the generated particles in the CMS detector is
simulated using the \GEANTfour
toolkit~\cite{Agostinelli:2002hh,Allison:2006ve}. During data
collection, an average of nine additional interactions occurred in
each bunch crossing (pileup). Pileup events are generated with \PYTHIA
and added to the generated hard-scattering events. The evolution of
beam conditions during data taking is taken into account by
reweighting the Monte Carlo (MC) simulation to match the distribution
of the number of pileup interactions observed in data.

\section{Event reconstruction and selection}
\label{eventReco}

The production of a \Z boson is identified through its decay into a
pair of isolated leptons (electrons or muons). Trigger selection
requires pairs of leptons with \pt exceeding predefined thresholds;
these thresholds were changed during the data acquisition period
because of the increasing instantaneous luminosity. For both lepton
types threshold pairs of 17\GeV and 8\GeV are used for most of the data
sample. The electron triggers include isolation requirements in order
to reduce the misidentification rate. Triggered events are
reconstructed using the particle-flow
algorithm~\cite{CMS-PAS-PFT-09-001,CMS-PAS-PFT-10-001}, which combines
the information from all CMS subdetectors to reconstruct and classify
muons, electrons, photons, charged hadrons, and neutral hadrons.

Electrons are selected with $\pt>20\GeV$ in the fiducial region of
pseudorapidity $\abs{\eta} < 2.4$, but excluding the region $1.44 <
\abs{\eta} < 1.57$ between the barrel and the end caps of ECAL to
ensure uniform quality of reconstruction. The electron identification
criteria~\cite{CMS-PAS-EGM-10-004,Chatrchyan:2013dga} comprise
requirements on the distance in $\eta$--$\phi$ space between the
cluster barycenter and the electron track extrapolation, where $\phi$
is the azimuthal angle measured in the plane transverse to the beams,
and the size and the shape of the electromagnetic shower in the
calorimeter. Electron-positron pairs consistent with photon conversion
are rejected.  Electron isolation is evaluated using all particles
reconstructed with the particle-flow algorithm within a cone around
the electron direction of radius $\DR=0.3$, where $\DR = \sqrt{\smash[b]{(\Delta
  \eta)^2 + (\Delta \phi)^2}}$ is the distance in the $\eta$--$\phi$
plane. An isolation variable is defined as
$I_{\text{rel}}=(I_{\text{charged}} +
I_{\text{photon}}+I_{\text{neutral}})/\pt^\Pe$,
where $I_{\text{charged}}$, $I_{\text{photon}}$,
$I_{\text{neutral}}$ are respectively the \pt sums of all charged
hadrons, photons, and neutral hadrons in the cone of interest, and
$\pt^\Pe$ is the electron transverse momentum. The
selection requires $I_{\text{rel}} < 0.15$.  Isolation variables are
sensitive to contamination from pileup events and thus a correction
for this effect is necessary for the high pileup environment of the
LHC collisions.  Only the particles consistent with originating from
the reconstructed primary vertex of the event, the vertex with the
largest quadratic sum of its constituent tracks' \pt, are included in
the calculation of $I_{\text{charged}}$. The $I_{\text{photon}}$
and $I_{\text{neutral}}$ components are corrected using the jet area
subtraction approach~\cite{Cacciari:2007fd}.

The selected muons must have $\pt > 20$\GeV and $\abs{\eta} < 2.4$.
Muon identification criteria are based on the quality of the global
track reconstruction, which includes both tracker and muon
detectors. Muons from cosmic rays are removed with requirements on the
impact parameter with respect to the primary vertex. In order to
evaluate the isolation, the variables $I_{\text{charged}}$,
$I_{\text{photon}}$, and $I_{\text{neutral}}$ are computed within
a cone of radius $\DR = 0.4$ around the trajectory of the muon
candidate, and $I_{\text{rel}}$ is required to be less than
0.2. Charged hadrons from pileup interactions are rejected by
requiring their tracks to be associated with the primary vertex. The
transverse momentum sum of the charged hadrons that are not associated
with the primary vertex is used to estimate the contribution from the
neutral particles produced in the pileup interactions; half of this
sum is subtracted from the isolation variable.

The two highest-\pt, same-flavor, oppositely charged, and isolated
leptons are selected to form the \Z-boson candidate if their invariant
mass lies between 71 and 111\GeV. The lepton pair is required to be
associated with the primary vertex of the event. Leptons associated
with the primary vertex and passing the isolation criteria are removed
from the collection of particles used for jet clustering.

For jet reconstruction, charged-particle tracks not associated with
the primary vertex are removed from the collection of particles used
for clustering. In this way, the dominant part of the pileup
contamination of the events of interest is suppressed. The remaining
particles are used as input to the jet clustering, which is based on
the anti-\kt algorithm~\cite{Cacciari:2008gp} as implemented in the
\textsc{FastJet} package~\cite{Cacciari:2005hq,Cacciari:2011ma}, with
a distance parameter in the rapidity-azimuth plane of 0.5. In order to
reject misreconstructed jets and instrumental noise, identification
quality criteria are imposed on the jets based on the energy fraction
of the charged, electromagnetic, and neutral hadronic components, and
requiring at least one charged particle in the jet.

Several effects contribute to bias the measured jet energy, compared
with the value it would acquire by clustering stable particles
originating from the fragmented hard-scattered partons and from the
underlying event. The sources of energy bias are pileup interactions,
detector noise, and detector response nonuniformities in $\eta$ and
nonlinearities in \pt.  The jet energy scale (JES)
calibration~\cite{Chatrchyan:1369486} relies on a combination of
\PYTHIA multijet simulations and measurements of exclusive dijet and
photon+jet events from data.  The corrections are parameterized in
terms of the uncorrected \pt and $\eta$ of the jet, and applied as
multiplicative factors scaling the four-momentum vector of each
jet. These factors include the correction for the contribution from
neutral pileup particles using the jet area
approach~\cite{Cacciari:2007fd}, and corrections for residual
discrepancies between data and simulation. The correction factors
range between 1.0 and 1.2, depend mostly on \pt, and are approximately
independent of $\eta$.

Furthermore, the jet energy resolution (JER) in data is known to be
worse than in the simulation, therefore the simulated resolution is
degraded to compensate for this effect. The difference between
the reconstructed jet transverse momentum and the corresponding generated
one is scaled in the simulation so as to reproduce the observed
resolution.

A minimum threshold of $\pt > 30$\GeV is required for the jets to
reduce contamination from the underlying event. Only jets with
$\abs{\eta} < 2.4$ are considered, and jets are required to be
separated from each lepton of the \Z candidate by $\DR\ge0.5$ in the
$\eta$--$\phi$ plane.

\section{Signal efficiency and background}
\label{sec::TaP}

A ``tag-and-probe'' technique~\cite{CMS:2011aa} is used to estimate
efficiencies for trigger selection, event reconstruction, and the
offline selection of the Z+jets sample. Scaling factors derived from
the ratio between the data and simulation efficiencies are used to
reweight simulated events in order to compensate for the residual
data-simulation differences. The correction is determined as a
function of \pt and $\eta$ of the leptons, and background components
are resolved using a binned extended maximum-likelihood fit of the
dilepton invariant-mass distribution between 60 and 120\GeV.  The
signal component of the distribution, which is taken from the
Drell--Yan simulated sample, is convolved with a Gaussian function to
account for the resolution difference between data and simulation.
The background contribution is modeled by an exponential function
multiplied by an error function describing the kinematic threshold due
to binning of the probe lepton \pt. The combined single-flavor
identification efficiency is the product of contributions from the
trigger, event reconstruction, and offline selection.  The same
technique is used on the data and in the simulation. The trigger
efficiency of the data ranges between 94\% and 99\% for electrons and
between 82\% and 97\% for muons.  The combined identification and
isolation efficiency depends on the \pt and $\eta$ of the leptons; it
ranges between 68\% and 91\% for electrons and between 86\% to 99\%
for muons.

The fiducial acceptance for muons and electrons is different, since
the latter are not well reconstructed in the transition region between
the barrel and end cap electromagnetic calorimeters.  In order to
facilitate the combination of results from the
$\cPZ\rightarrow\Pep\Pem$ and $\cPZ\rightarrow\Pgmp\Pgmm$ final
states, this difference is evaluated using the simulation, giving a
correction to the \EE cross section, applied within the unfolding
procedure described in the next section, that amounts to 8\%.

Several background processes can produce or mimic two reconstructed
opposite-sign same-flavor leptons. The largest contribution comes from
\ttbar production , while diboson production contribute near the \Z-boson invariant-mass
peak. Other minor contributions arise from
$\cPZ\rightarrow\Pgt^+\Pgt^-$ as well as single-top and \PW+jets
events. The contamination from multijet events produced through the
strong interaction is negligible, as established with a control sample
in which the two leptons in each event have the same
charge~\cite{Chatrchyan:2013tna}.  The total
contribution of the backgrounds is approximately 1\% of the total
yield of the selected events, and it increases as a function of jet
multiplicity. At the highest measured jet multiplicities it reaches
values up to 10\%. The background subtraction procedure is performed after scaling
the number of background events to the integrated luminosity in the
data sample using the corresponding cross section for each background
process.

\begin{figure}[hbtp]
\centering
\includegraphics[width=0.49\textwidth]{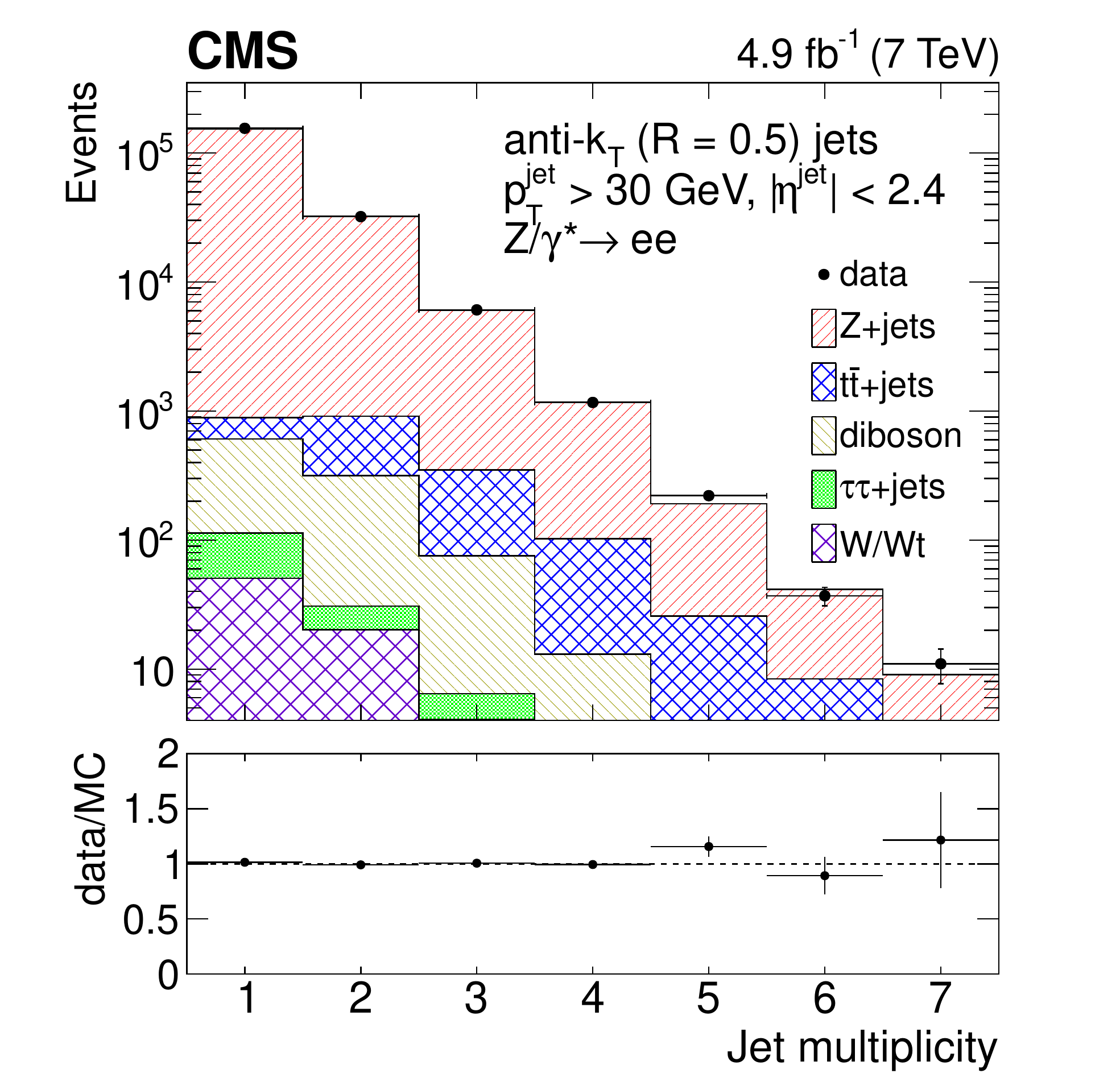}
\includegraphics[width=0.49\textwidth]{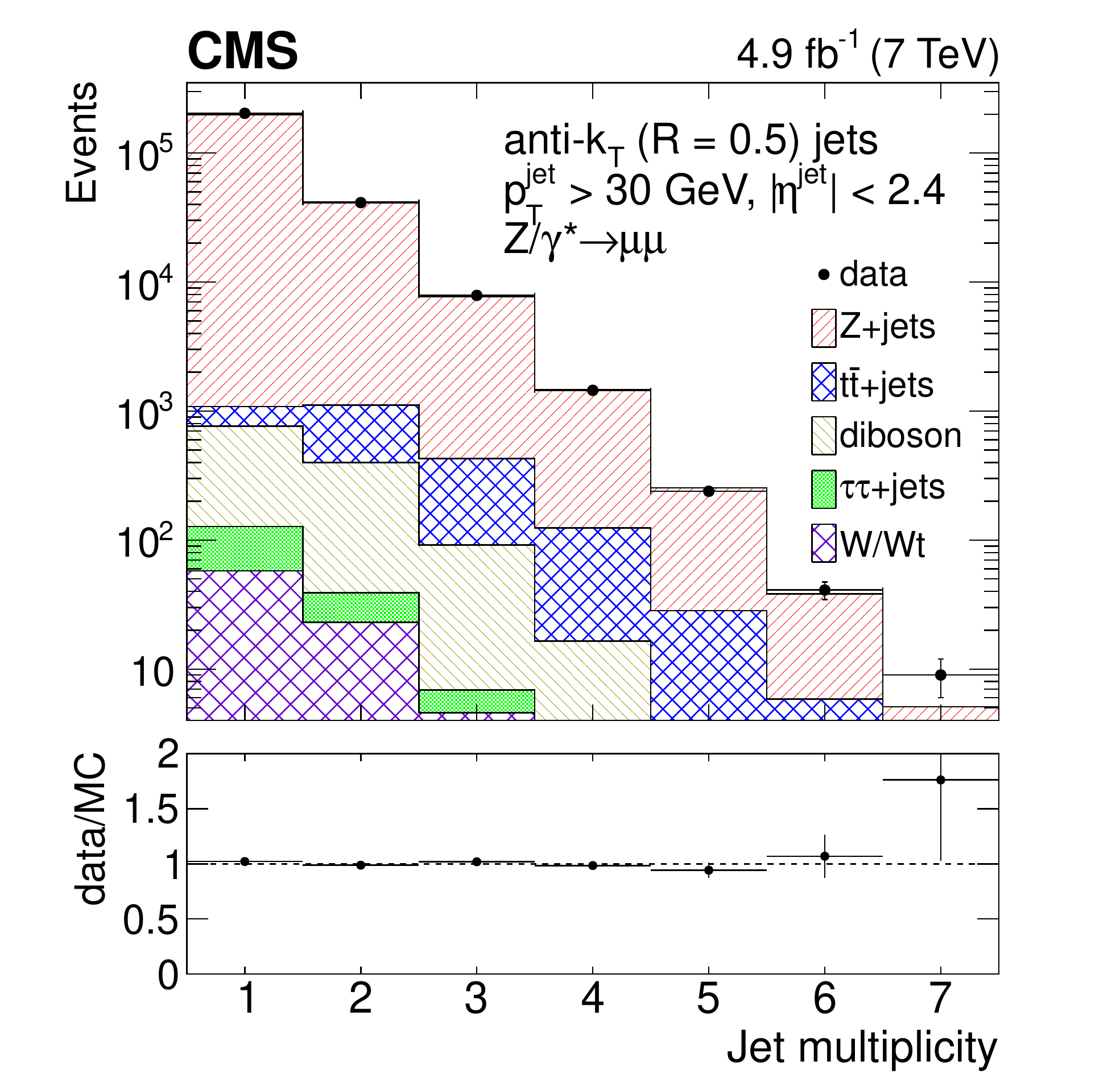}
\caption{Distributions of the exclusive jet multiplicity for the
  electron channel (\cmsLeft) and muon channel (\cmsRight). Data are
  compared to the simulation, which is the sum of signal and
  background events. Scale factors have been used to correct
  simulation distributions for residual efficiency differences with
  respect to data. No unfolding procedure is applied. Only statistical
  uncertainties are shown.}
\label{fig:invMassCombined}

\end{figure}

The exclusive jet multiplicity in the selected events is shown in
Fig.~\ref{fig:invMassCombined}. For both leptonic decay channels, the
data show overall agreement with combined signal and background
samples from the simulation. The ratio between the cross sections as a
function of jet multiplicity in data and in signal plus
background simulation, shown in the bottom part of the figure, is
compatible with unity within the uncertainties.

\section{Unfolding}
\label{sec:unfolding}

The distributions of the observables are corrected for event selection
efficiencies and for detector resolution effects back to the stable particle
level, in order to compare with predictions from event generators
simulating \cPZ+jets final states. Particles are considered stable if
their proper average lifetime $\Pgt$ satisfies $c\Pgt > 10$\cm.  The
correction procedure is based on unfolding techniques, as implemented
in the \textsc{RooUnfold} toolkit~\cite{arxiv:1105.1160}, which
provides both the ``singular value decomposition'' (SVD)
method~\cite{nima:a372_469_481} and the iterative algorithm based on the
Bayes' theorem~\cite{nima:a362_487_498}.  Both algorithms use a
``response matrix'' that correlates the values of the observable with
and without detector effects. 

The response matrix is evaluated using \cPZ+jets events, generated by
\MADGRAPH followed by \PYTHIA, with full detector simulation. For
generator-level events, leptons and jets are reconstructed from the
collection of all stable final-state particles using criteria that
mimic the reconstructed data.  Electrons and muons with the highest
\pt above 20\GeV in the pseudorapidity range $\abs{\eta} < 2.4$ are
selected as \Z-boson decay products. In order to include the effects
of final-state electromagnetic radiation in the generator-level
distributions, the electron and muon candidates are reconstructed by
clustering the leptons with all photons in a cone of radius $\DR =
0.1$ in the $\eta$--$\phi$ plane.  Leptons from \Z-boson decay are
removed from the particle collection used for the jet clustering at
generator level. The remaining particles, excluding neutrinos, are
clustered into jets using the anti-\kt algorithm.  A generated jet is
included in the analysis if it satisfies $\pt > 30$\GeV, $\abs{\eta} <
2.4$; the jet must contain at least one charged particle, to match the
jet reconstruction quality requirements used for data analysis, and
the distance of the jet from the leptons forming the \Z-boson
candidate is larger than $\DR = 0.5$.

The unfolded distributions are obtained with the SVD algorithm. As a
cross-check, the unfolding of the distributions is also performed with
the D'Agostini method, which leads to compatible results within
statistical uncertainties. The unfolding has a small effect on the jet
$\eta$ distributions, with migrations among the bins of a few percent
for central jets and up to 10\% in the outer regions. Larger unfolding
effects are observed in the other distributions: up to 20\% for the
jet multiplicity, between 10\% and 20\% for the jet \pt, and between
10\% and 30\% for the \HT distribution.

\section{Systematic uncertainties}
\label{systematics}

The sources of systematic uncertainties that affect the \cPZ+jets
cross section measurement are divided into the following
categories: jet energy scale (JES) and jet energy resolution
(JER)~\cite{Chatrchyan:1369486}, unfolding procedure, efficiency
correction and background subtraction, pileup reweighting procedure,
and integrated luminosity measurement.

Jet energy scale and resolution uncertainties affect the jet \pt
reconstruction and the determination of \HT. Each JES correction
factor has an associated uncertainty that is a function of the $\eta$
and \pt of the jet. The difference in the distribution of an
observable, after varying the JES both up and down by one standard
deviation, is used as an estimate of the JES systematic
uncertainty. Similarly, the effect of the systematic uncertainties in
the scaling factor used in the JER degradation is estimated by  
varying its value up and down by one standard deviation.

The uncertainty in the unfolding procedure is due to both the
statistical uncertainty in the response matrix from the finite size of
the simulated sample and to any dependence on the signal model
provided by different event generators. The statistical uncertainty is
computed using a MC simulation, which produces variants of the matrix
according to random Poisson fluctuations of the bin contents. The
entire unfolding procedure is repeated for each variant, and the
standard deviation of the obtained results is used as an estimate of
this uncertainty. The systematic uncertainty due to the generator
model is estimated from the difference between events simulated with
\MADGRAPH and \SHERPA at detector response level.  The overall
unfolding uncertainty is taken to be either the statistical
uncertainty alone, in the case where the results from the two event
generators agree within one standard deviation, or the sum in
quadrature of the simulation statistical uncertainty and the
difference between the two MC generators.

Additional uncertainty arises from the efficiency corrections and from
the background subtraction. The contribution due to efficiency
corrections is estimated by adding and subtracting the statistical
uncertainties from the tag-and-probe fits.  The systematic uncertainty
from the background subtraction procedure is small relative to the
other sources. For \ttbar and diboson processes, the uncertainty in
the normalization arises from both the theoretical uncertainty in the
inclusive cross section and the difference between the theoretical
prediction of the cross section (as in Section~\ref{data&mc}) and the
corresponding CMS
measurement~\cite{Chatrchyan:2012bra,Chatrchyan:2012ria,Chatrchyan:2012bd,Chatrchyan:2012sga}.
The largest of these two values is taken as the magnitude of the
uncertainty. As observed in previous
studies~\cite{Chatrchyan:2013tna}, the single top quark and \PW+jets
contributions are at the sub-per-mil level, and they are assigned a
100\% uncertainty.

Since the background contribution as a function of the jet
multiplicity is theoretically less well known than the fully inclusive
cross section, control data samples are used to validate the
simulation of this dependence. The modeling of the dominant \ttbar
background as a function of the jet multiplicity is compared with the
data using a control sample enriched in \ttbar events. This sample is
selected by requiring the presence of two leptons of different
flavors, \ie, $\Pe\Pgm$ combinations, and an agreement is found
between data and simulation at the 6\%
level~\cite{Chatrchyan:2013tna}. The CMS measurement of the \ttbar
differential cross section~\cite{Chatrchyan:2012saa}, using an event
selection compatible with the study presented in this paper, leads to
a production rate for events with six jets in simulation overestimated
by about 30\%. This difference is used as the estimated uncertainty
for the six-jets subsample.  Variations in the \MADGRAPH prediction
for \ttbar production from a change in the renormalization,
factorization, and matching scales, as well as from the PDF choice,
show that data and simulation agree within the estimated uncertainties.

The systematic uncertainty of the pileup reweighting procedure in MC
simulation is due to the uncertainties in the minimum-bias cross
section and in the instantaneous luminosity of the data sample. This
uncertainty is evaluated by varying the number of simulated pileup
interactions by ${\pm}5$\%. The measurement of the integrated
luminosity has an associated uncertainty of 2.2\% that directly
propagates to any cross section measurement.

The systematic uncertainties (excluding luminosity) used for the
combination of the electron and muon samples are summarized in
Tables~\ref{tab:systPt},~\ref{tab:systEta}, and~\ref{tab:systHt}.
\begin{table*}[htbH]
\topcaption{Sources of uncertainties (in percent) in the differential
  exclusive cross section and in the differential cross sections as a
  function of the jet \pt, for each of the four highest \pt jets
  exclusively. The constant luminosity uncertainty is not included in
  the total.}
\centering
\begin{scotch}{lccccc}
Systematic uncertainty (\%)& $\sigma(\cPZ/\gamma^{\ast}+\text{jets})$&$\frac{\rd\sigma}{\rd\pt}$($1^{\mathrm{st}}$ jet)&$\frac{\rd\sigma}{d\pt}$($2^{\mathrm{nd}}$ jet)&$\frac{\rd\sigma}{\rd\pt}$($3^{\mathrm{rd}}$ jet)&$\frac{\rd\sigma}{\rd\pt}$($4^{\mathrm{th}}$ jet)\\
\hline
JES+JER                         & 2.0--18    & 4.9--8.7      & 6.3--16      & 8.8--15    & 15--23  \\
Unfolding                       & 1.7--9.2   & 1.3--22       & 0.5--21      & 0.8--13    & 0.3--12  \\
Efficiency                      & 0.3       & 0.3          & 0.3         & 0.3       & 0.3  \\
Background                      & 0.1--25    & 0.1--0.4      & 0.6--1.8     & 0.6--1.0   & 0.9--1.5  \\
Pileup                          & 0.3--0.8   & 0.2--2.7      & 0.3--0.6     & 0.2--0.7   & 0.4--1.0  \\
\hline
Total syst. uncertainty (\%)        & 2.7--32    & 5.1--24       & 9.0--27      & 10--20   & 17--23  \\
\hline
Statistical uncertainty (\%)  & 0.7--6.4   & 0.1--7.2      & 1.4--12      & 3.0--13    & 4.3--19  \\
\end{scotch}

\label{tab:systPt}
\end{table*}

\begin{table*}[htbH]
\topcaption{Sources of uncertainties (in percent) in the differential
  cross sections as a function of $\eta$, for each of the four highest
  \pt jets exclusively. The constant luminosity uncertainty is not
  included in the total.}
\centering
\begin{scotch}{lcccc}
Systematic uncertainty (\%) &  $\frac{\rd\sigma}{\rd\eta}$ ($1^{\mathrm{st}}$ jet)& $\frac{\rd\sigma}{\rd\eta}$ ($2^{\mathrm{nd}}$ jet)& $\frac{\rd\sigma}{\rd\eta}$ ($3^{\mathrm{rd}}$ jet)& $\frac{\rd\sigma}{\rd\eta}$ ($4^{\mathrm{th}}$ jet)\\
\hline
JES+JER           & 3.5--8.2   & 7.2--8.9     & 9.4--12   & 13--15  \\
Unfolding         & 6.5--13    & 8.4--11      & 5.0--12   & 6.4--13  \\
Efficiency        & 0.3       & 0.3         & 0.3      & 0.3  \\
Background        & 0.2       & 0.3--0.5     & 0.6--1.1  & 0.9--1.0  \\
Pileup            & 0.2--0.4   & 0.3--0.5     & 0.3--0.7  & 0.5--1.2  \\
\hline
Total syst. uncertainty (\%) & 7.8--17  & 11--15  & 11--19  & 15--23  \\
\hline
Statistical uncertainty (\%)        & 0.6--1.0   & 0.9--1.4     & 2.4--3.6    & 7.6--12  \\
\end{scotch}

\label{tab:systEta}
\end{table*}

\begin{table*}[htbH]
\topcaption{Sources of uncertainties (in percent) in the differential
  cross sections as a function of \HT and inclusive jet
  multiplicity. The constant luminosity uncertainty is not included
  in the total.}
\centering
\begin{scotch}{lcccc}

Systematic uncertainty (\%) & $\frac{\rd\sigma}{\rd\HT}$,
$N_{\text{jet}}\geq 1$ & $\frac{\rd\sigma}{\rd\HT}$,
$N_{\text{jet}}\geq 2$ &$\frac{\rd\sigma}{\rd\HT}$,
$N_{\text{jet}}\geq 3$ & $\frac{\rd\sigma}{\rd\HT}$,
$N_{\text{jet}}\geq 4$ \\
\hline
JES+JER           & 4.5--9.1     & 7.0--11   & 8.6--13    & 11--17  \\
Unfolding         & 0.4--17      & 2.1--18   & 3.1--22    & 4.9--23  \\
Efficiency        & 0.2--0.3     & 0.3      & 0.3--0.4   & 0.3  \\
Background        & 0.1--0.7     & 0.3--0.7  & 0.5--0.8   & 0.6--1.1  \\
Pileup            & 0.1--2.3     & 0.1--2.2  & 0.3--1.0   & 0.5--1.0  \\
\hline
Total syst. uncertainty (\%) & 4.6--19    & 7.8--21   & 10--26   & 12--25  \\
\hline
Statistical uncertainty (\%)         & 0.6--4.1     & 0.9--3.3    & 2.3--5.6    & 8.6--17  \\
\end{scotch}

\label{tab:systHt}
\end{table*}

\section{Results and comparison with theoretical predictions}

The results presented for observable quantities are obtained by
combining the unfolded distributions for both leptonic channels into
an uncertainty-weighted average for a single lepton
flavor. Correlations between systematic uncertainties for the electron
and muon channels are taken into account in the combination.  Fiducial
cross sections are shown, without further corrections for the
geometrical acceptance or kinematic selection, for leptons and jets.
All the results are compared with theoretical distributions, produced
with the \RIVET toolkit~\cite{Buckley:2010ar}, obtained
with the generator-level phase space definition and on final-state
stable particles as discussed in
Section~\ref{sec:unfolding}. Neutrinos are excluded from the
collection of stable particles.

Theoretical predictions at leading order in pQCD are computed with the
\MADGRAPH~5.1.1 generator followed by \PYTHIA~6.424 with the Z2 tune
and CTEQ6L1 PDF set for fragmentation and parton shower
simulation. For the \MADGRAPH simulation, the factorization and
renormalization scales are chosen on an event-by-event basis as the
transverse mass of the event, clustered with the \kt algorithm down to
a 2$\rightarrow$2 topology, and $\kt$ at each vertex splitting,
respectively~\cite{Alwall:2008qv,Alwall:2007fs}.  The \MADGRAPH
predictions are rescaled to the available NNLO inclusive cross
section~\cite{Melnikov:2006kv}, which has a uniform associated
uncertainty of about 5\% that is not propagated into the figures.

Predictions at next-to-leading order in QCD are provided by
\SHERPA~2.$\beta$2~\cite{Gleisberg:2008fv,Schumann:2007mg,Gleisberg:2008ta,Hoeche:2009rj,Hoeche:2012yf},
using the CT10 NLO PDF set~\cite{CT10:2010}, in a configuration where
NLO calculations for \cPZ+0 and \cPZ+1~jet event topologies are merged
with leading-order matrix elements for final states with up to four
real emissions and matched to the parton shower. The NLO virtual
corrections are computed using the \BLACKHAT
library~\cite{Berger:2010gf}. In this calculation, the factorization
and renormalization scales are defined for each event by clustering
the 2$\rightarrow$$n$ parton level kinematics onto a core
2$\rightarrow$2 configuration using a \kt-type algorithm, and using
the smallest invariant mass or virtuality in the core configuration as
the scale~\cite{Hoeche:2012yf}. The default configuration for the
underlying event and fragmentation tune is used.

The third theoretical prediction considered is the NLO QCD calculation
for the \cPZ+1~jet matrix element as provided by the \POWHEGBOX
package~\cite{Nason:2004rx,Frixione:2007vw,Alioli:2010xd,Alioli:2010qp},
with CT10 NLO PDF set, and matched with the \PYTHIA parton shower
evolution using the Z2 tune. In this case, the factorization and
renormalization scales in the inclusive cross section calculation are
defined on an event-by-event basis as the \Z-boson \pt, while for the
generation of the radiation they are given by the \pt of the produced
radiation.

The comparison of these predictions with the corrected data are
presented in Figs.~\ref{fig:DiffMulti}--\ref{FinalJetHT}.  The effect
of PDF choice is shown in
Figs.~\ref{fig:DiffMultiPDF}--\ref{FinalJetHTPDF}.  The error bars on
the plotted data points represent the statistical uncertainty, while
cross-hatched bands represent the total experimental uncertainty
(statistical and systematic uncertainties summed in quadrature) after
the unfolding procedure. Uncertainties in the theoretical predictions
are shown in the ratio of data to simulation only. For the NLO
prediction, theoretical uncertainties are evaluated by varying
simultaneously the factorization and renormalization scales up and
down by a factor of two (for \SHERPA and \POWHEG). For the \SHERPA
prediction only, the resummation scale is changed up and down by a
factor $\sqrt{2}$ and the parton shower matching scale is changed by
10\GeV in both directions. The effect of the PDF choice is shown for
\SHERPA, by comparing the results based on CT10 PDF set with those
based on the alternative NLO PDFs MSTW2008 and
NNPDF2.1~\cite{Ball:2011mu}. The theoretical part of the plotted
uncertainty band for each PDF choice includes both the intrinsic PDF 
uncertainty, evaluated according to the prescriptions of the authors
of each PDF set, and the effect of the variation of $\pm 0.002$ in the
value of the strong coupling constant $\alpha_{\mathrm{S}}$ around the
central value used in the PDF.

\subsection{Jet multiplicity}

\begin{figure*}[hbtp]
\centering
\includegraphics[width=\cmsFigWidth]{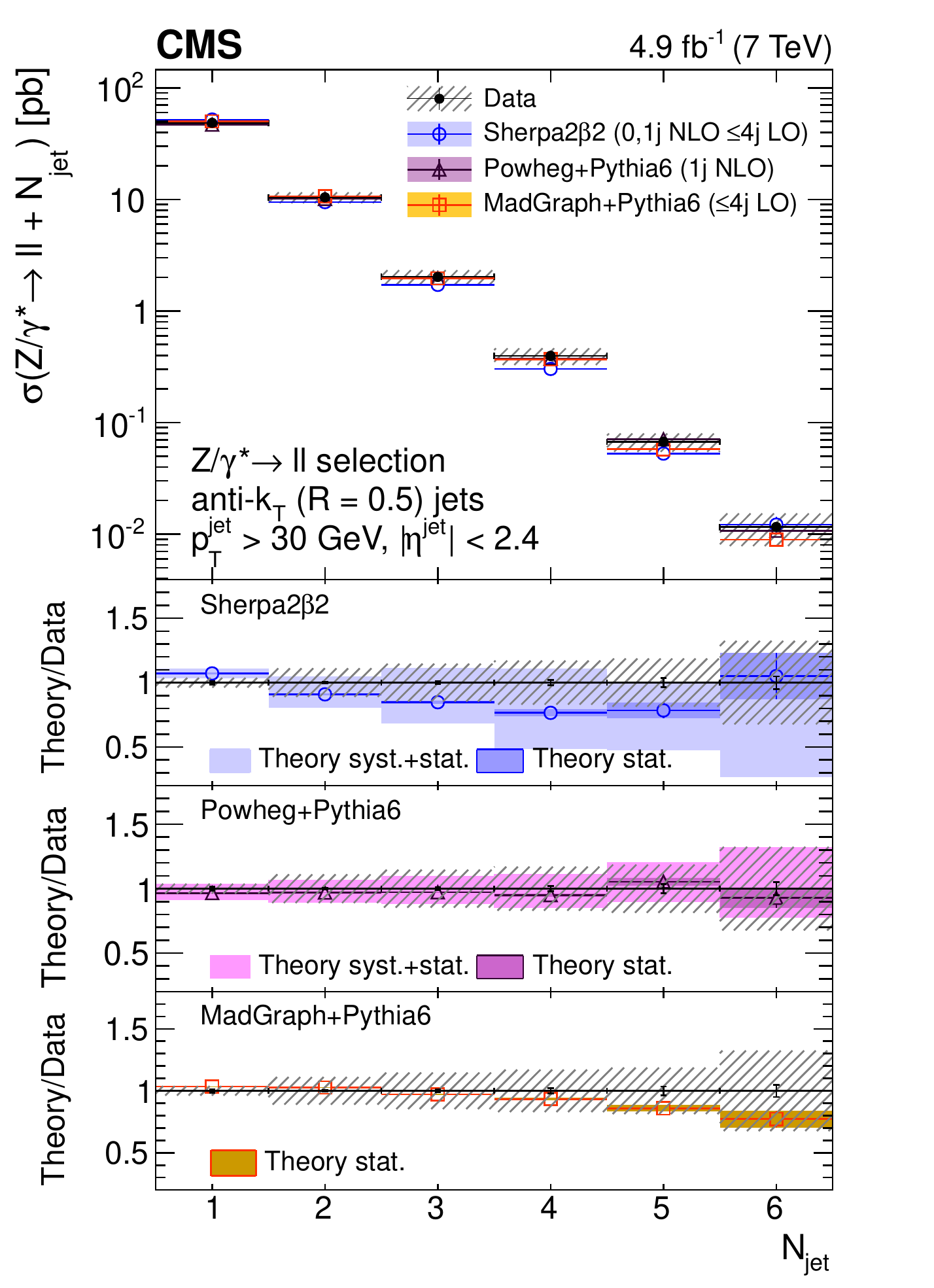}
\includegraphics[width=\cmsFigWidth]{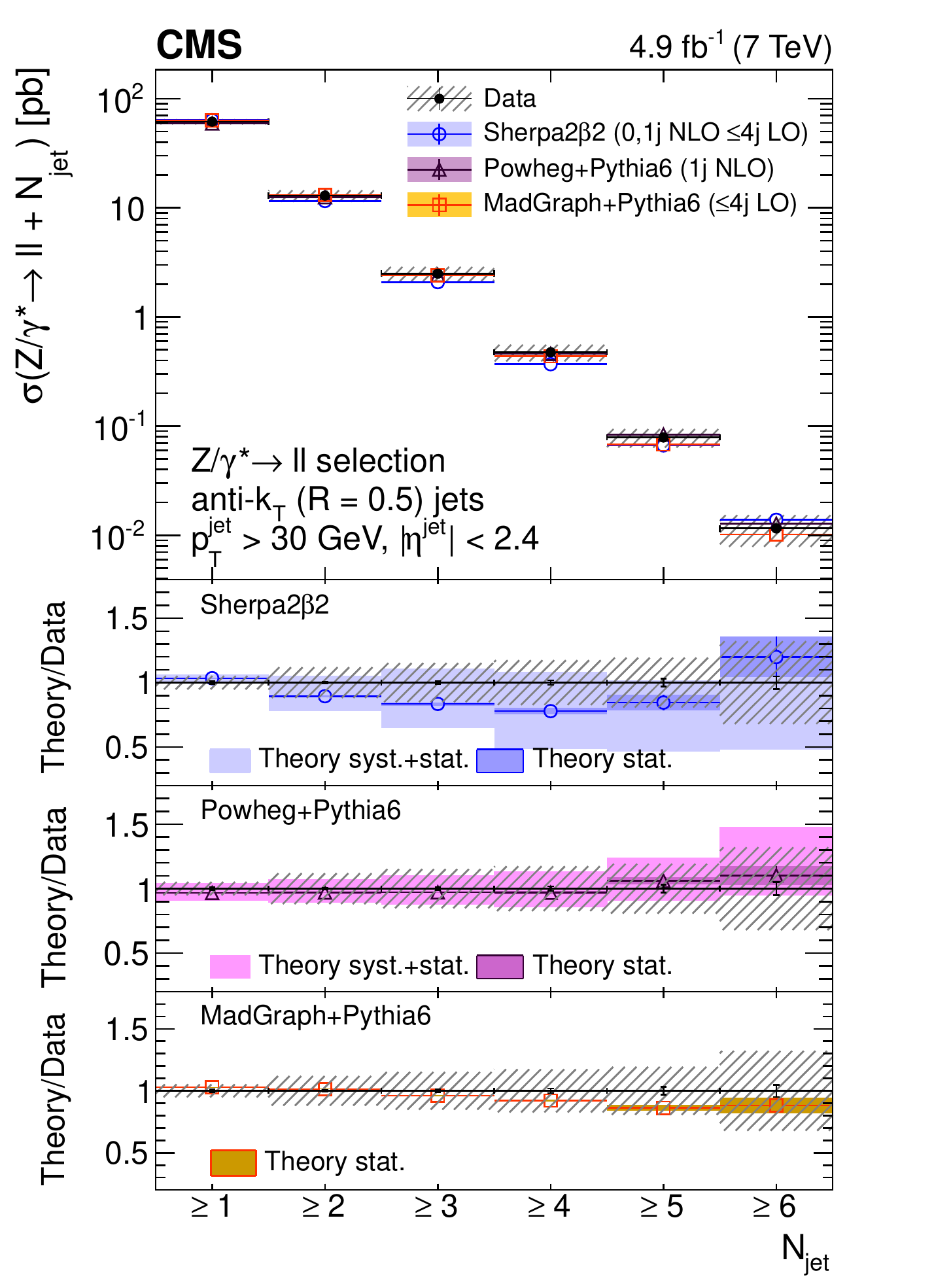}
\caption{Exclusive~(left) and inclusive~(right) jet
  multiplicity distributions, after the unfolding procedure,
  compared with \SHERPA, \POWHEG, and \MADGRAPH predictions. Error
  bars around the experimental points represent the statistical
  uncertainty, while cross-hatched bands represent statistical plus
  systematic uncertainty. The bands around theory predictions
  correspond to the statistical uncertainty of the generated sample
  and, for NLO calculations, to its combination with the systematic
  uncertainty related to scale variations.}
\label{fig:DiffMulti}

\end{figure*}

Figure~\ref{fig:DiffMulti} shows the measured cross sections as a
function of the exclusive and inclusive jet multiplicities, for a
total number of up to six jets in the final state. Beyond the sixth
jet, the measurement is not performed due to the statistical
limitation of the data and simulated samples. The trend of the jet
multiplicity represents the expectation of the pQCD prediction for a
staircase-like scaling, with an approximately constant ratio between
cross sections for successive
multiplicities~\cite{Berends:1989cf}. This result confirms the
previous observation, which was based on a more statistically limited
sample~\cite{Chatrchyan:2011ne}. Within the uncertainties, there is
agreement between theory and measurement for both the inclusive and
the exclusive distributions.

\subsection{Differential cross sections}

\begin{figure*}[hbtp]
\centering
\includegraphics[width=\cmsFigWidth]{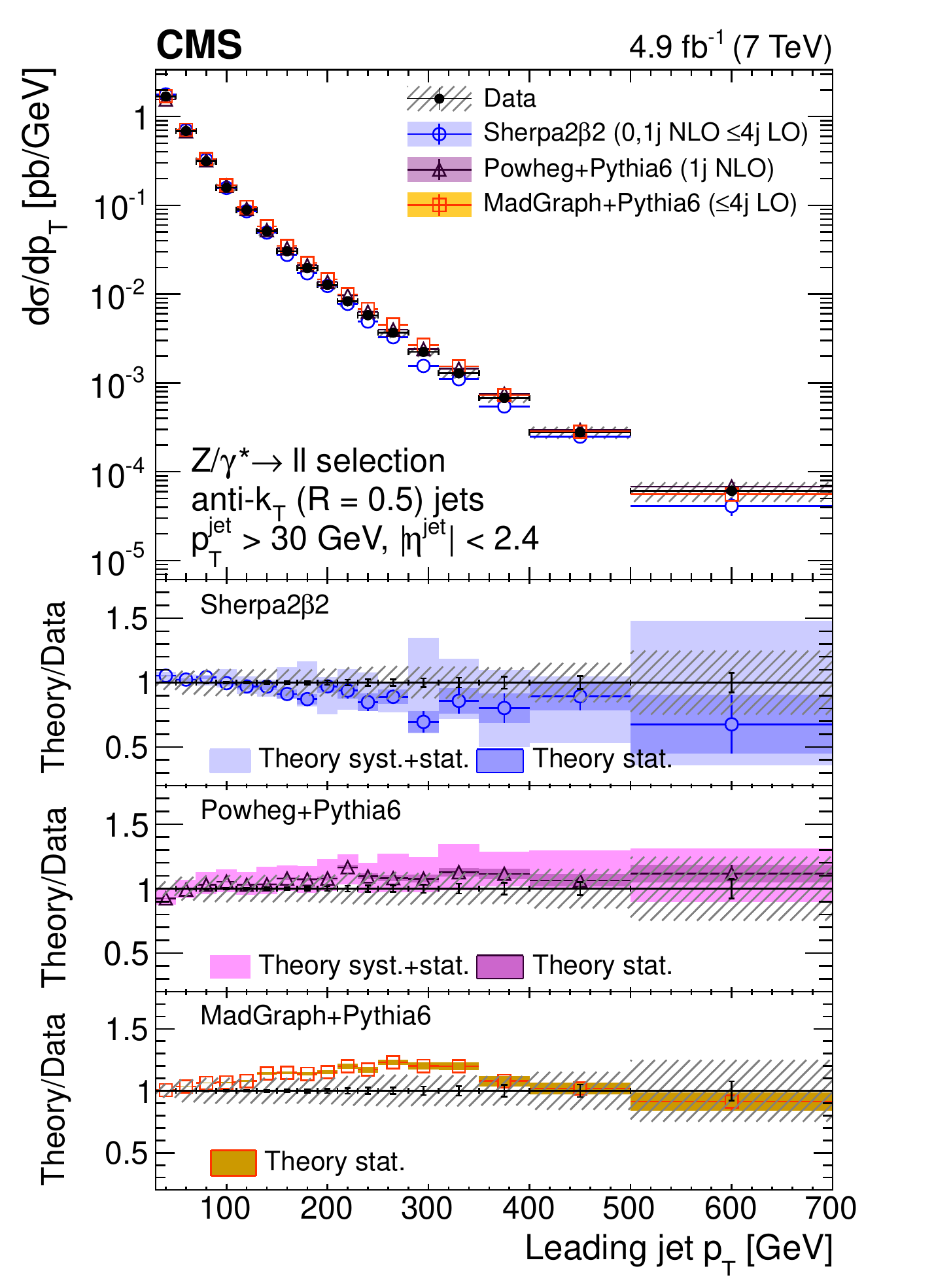}
\includegraphics[width=\cmsFigWidth]{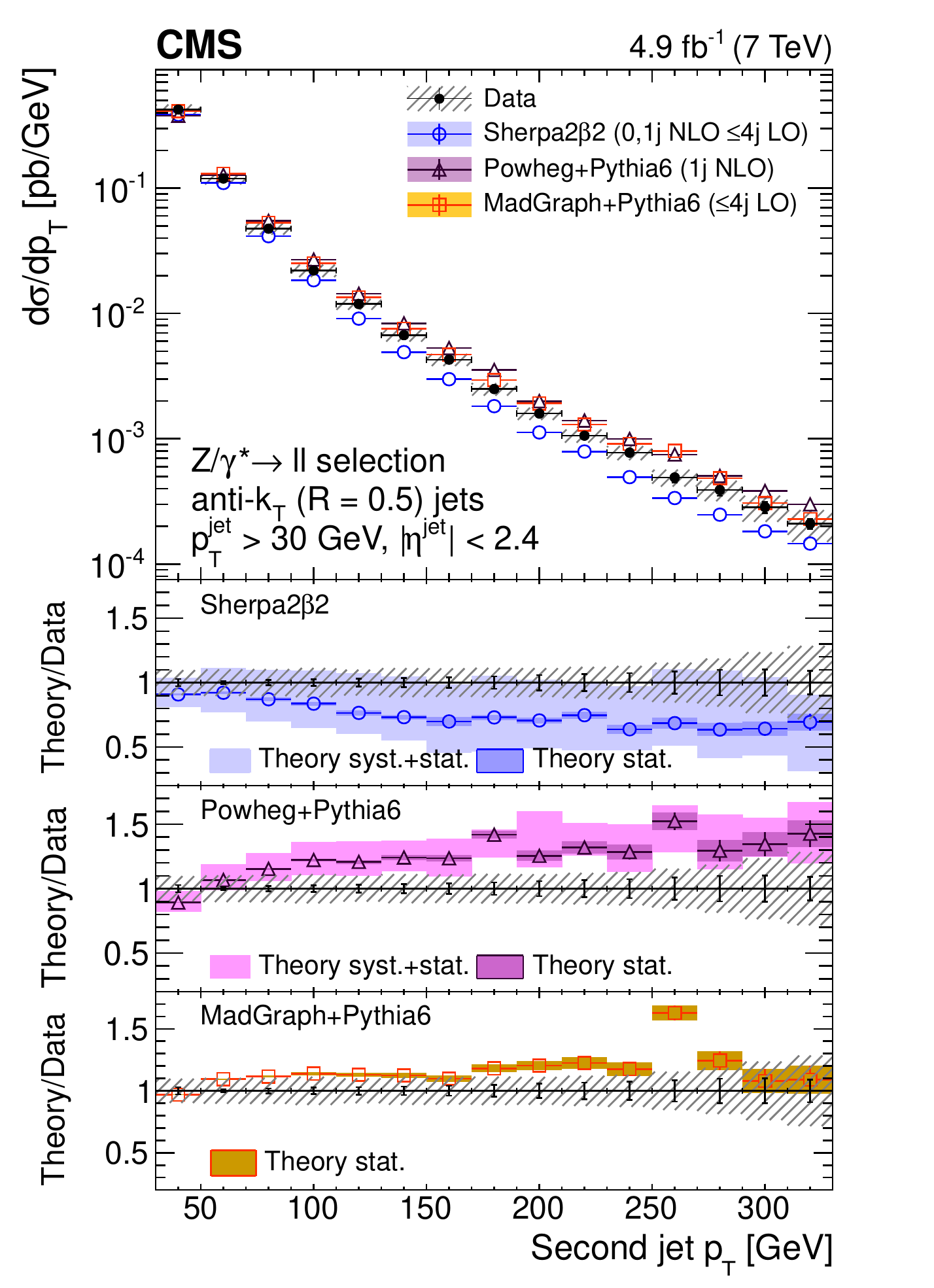}\\
\includegraphics[width=\cmsFigWidth]{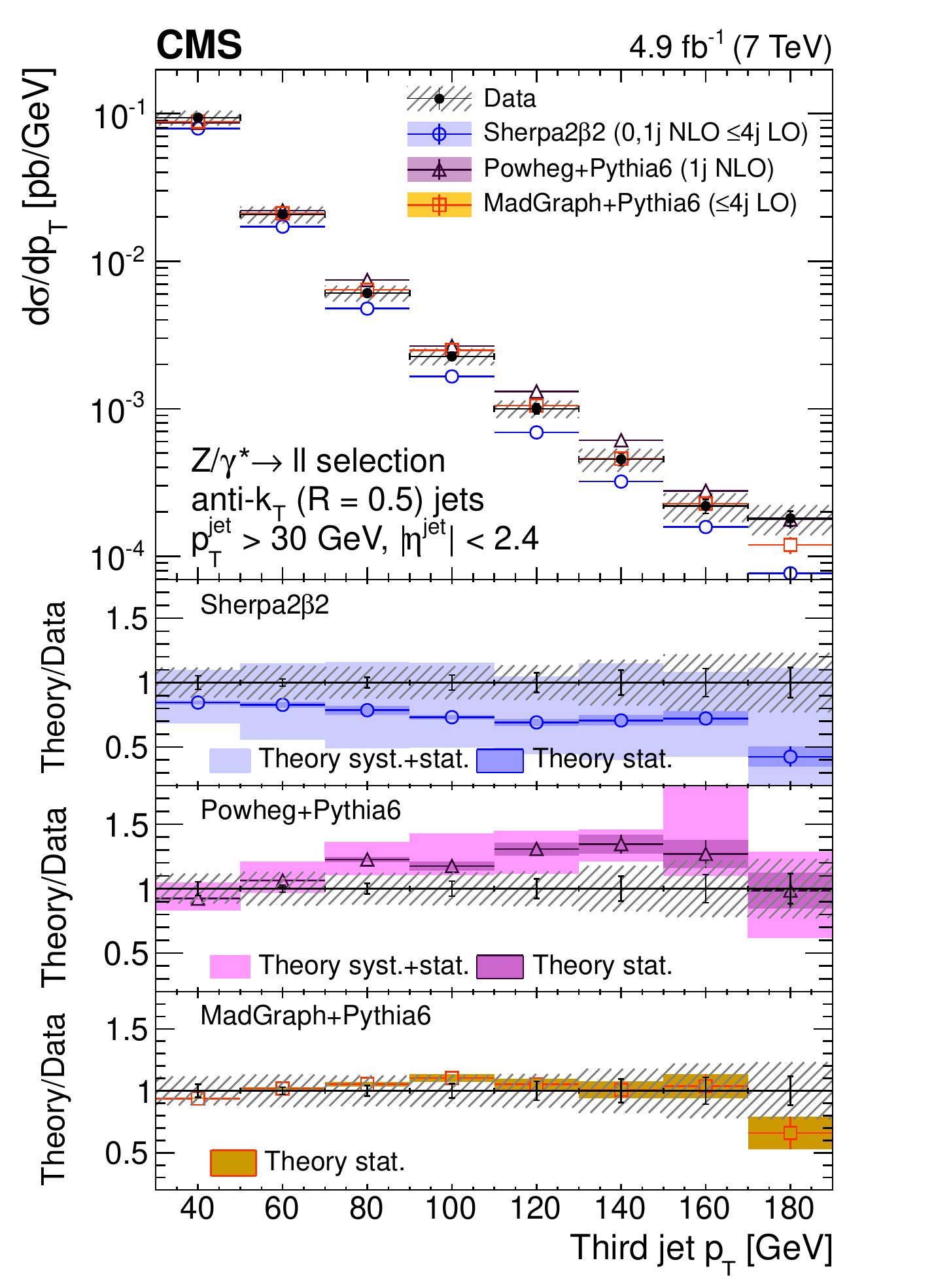}
\includegraphics[width=\cmsFigWidth]{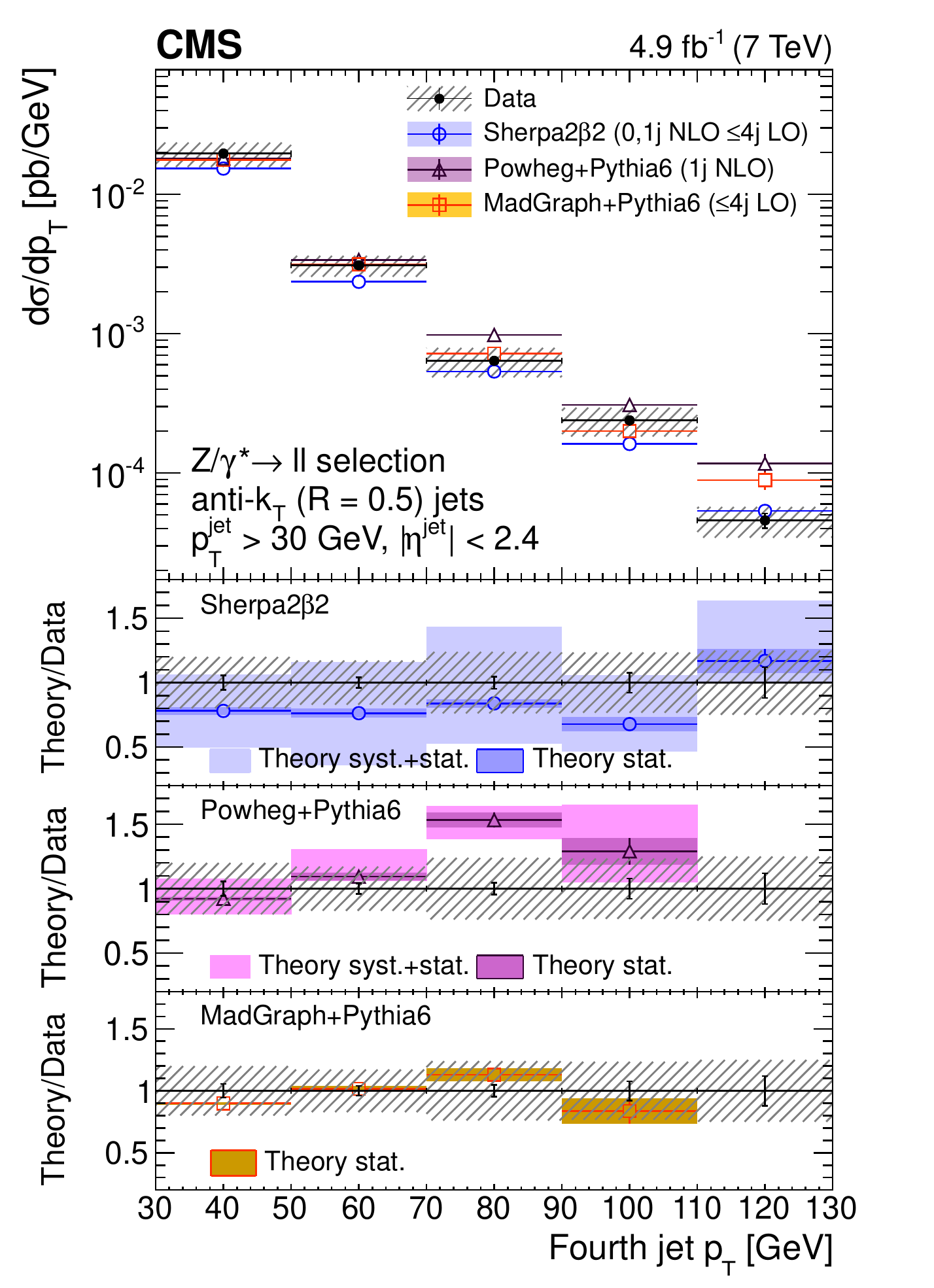}
\caption{Unfolded differential cross section as a function of \pt
  for the first (top left), second (top right), third (bottom left),
  and fourth (bottom right) highest \pt jets, compared with \SHERPA,
  \POWHEG, and \MADGRAPH predictions. Error bars around the
  experimental points represent the statistical uncertainty, while
  cross-hatched bands represent statistical plus systematic
  uncertainty. The bands around theory predictions correspond to the
  statistical uncertainty of the generated sample and, for NLO
  calculations, to its combination with systematic uncertainty related to
  scale variations.}
\label{FinalJetPt}

\end{figure*}

\begin{figure*}[hbtp]
\centering
\includegraphics[width=\cmsFigWidth]{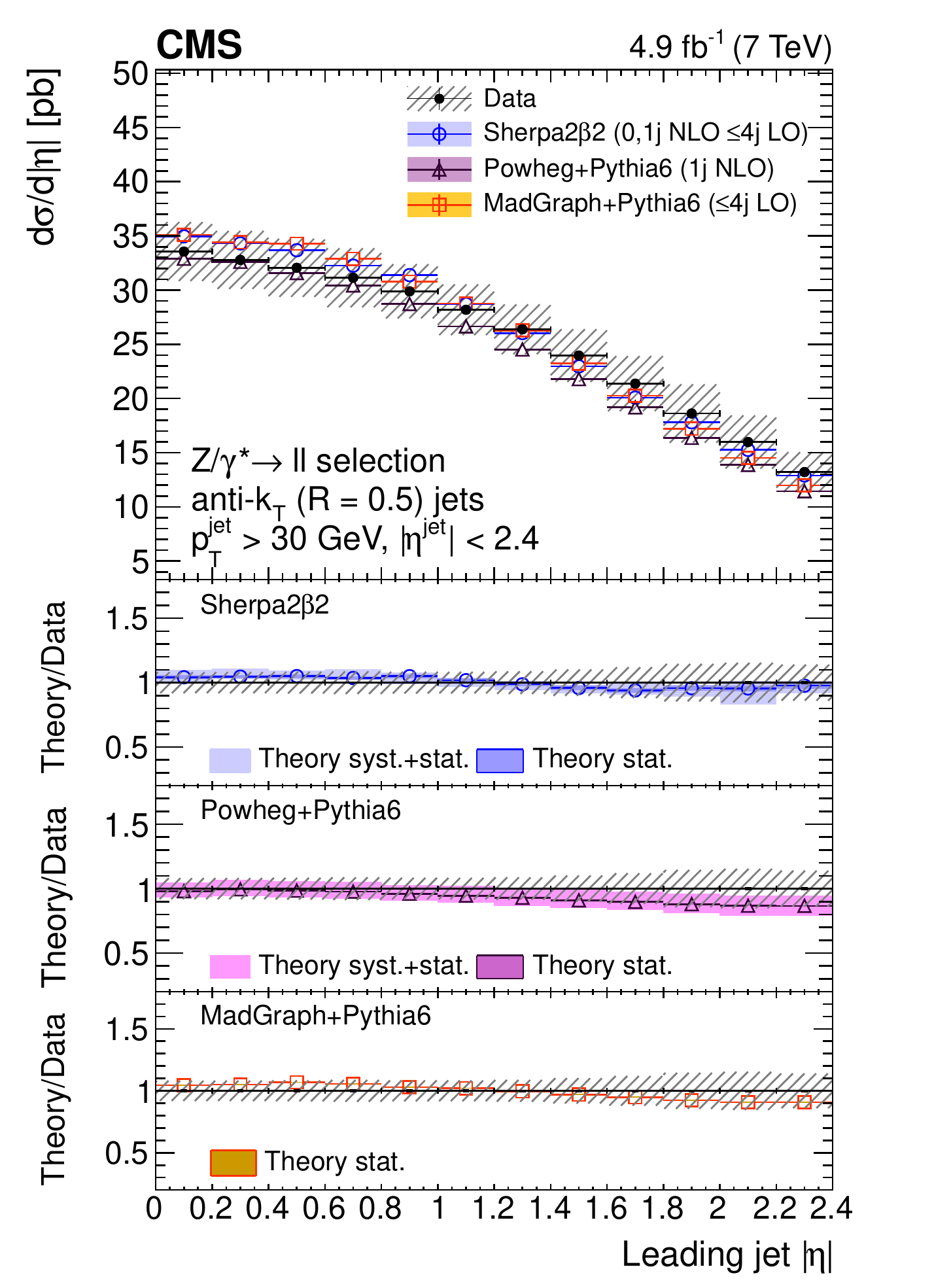}
\includegraphics[width=\cmsFigWidth]{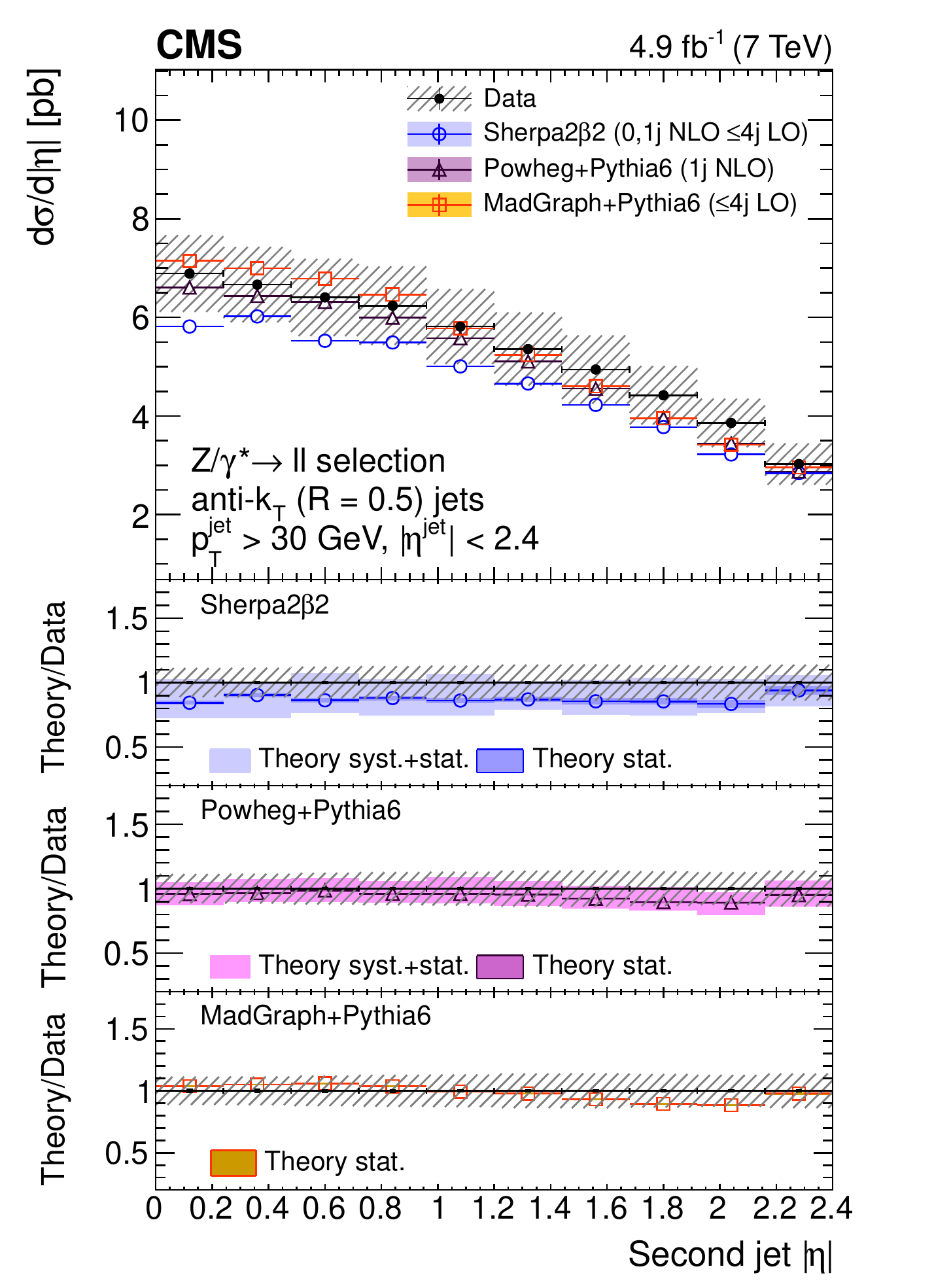}\\
\includegraphics[width=\cmsFigWidth]{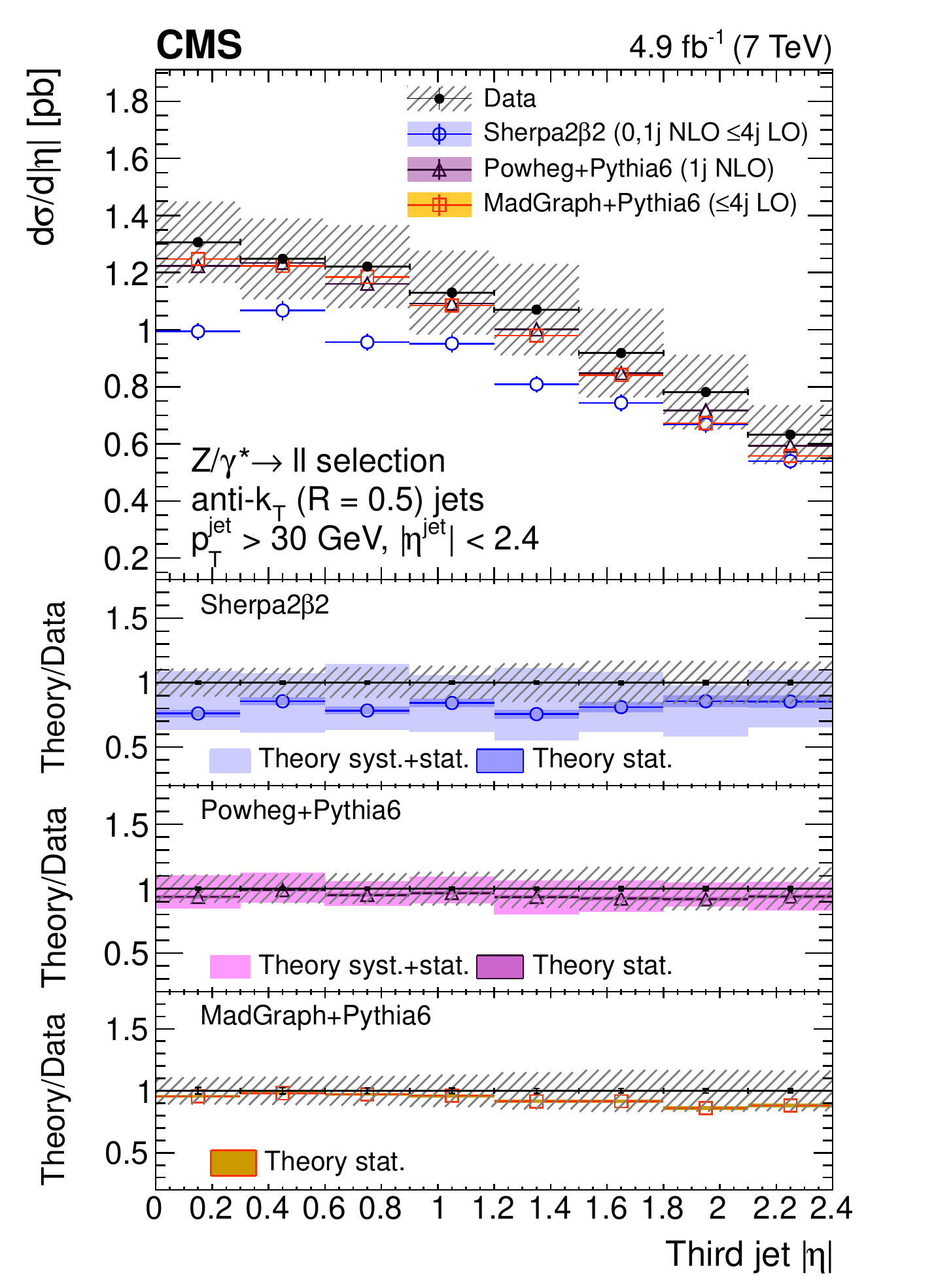}
\includegraphics[width=\cmsFigWidth]{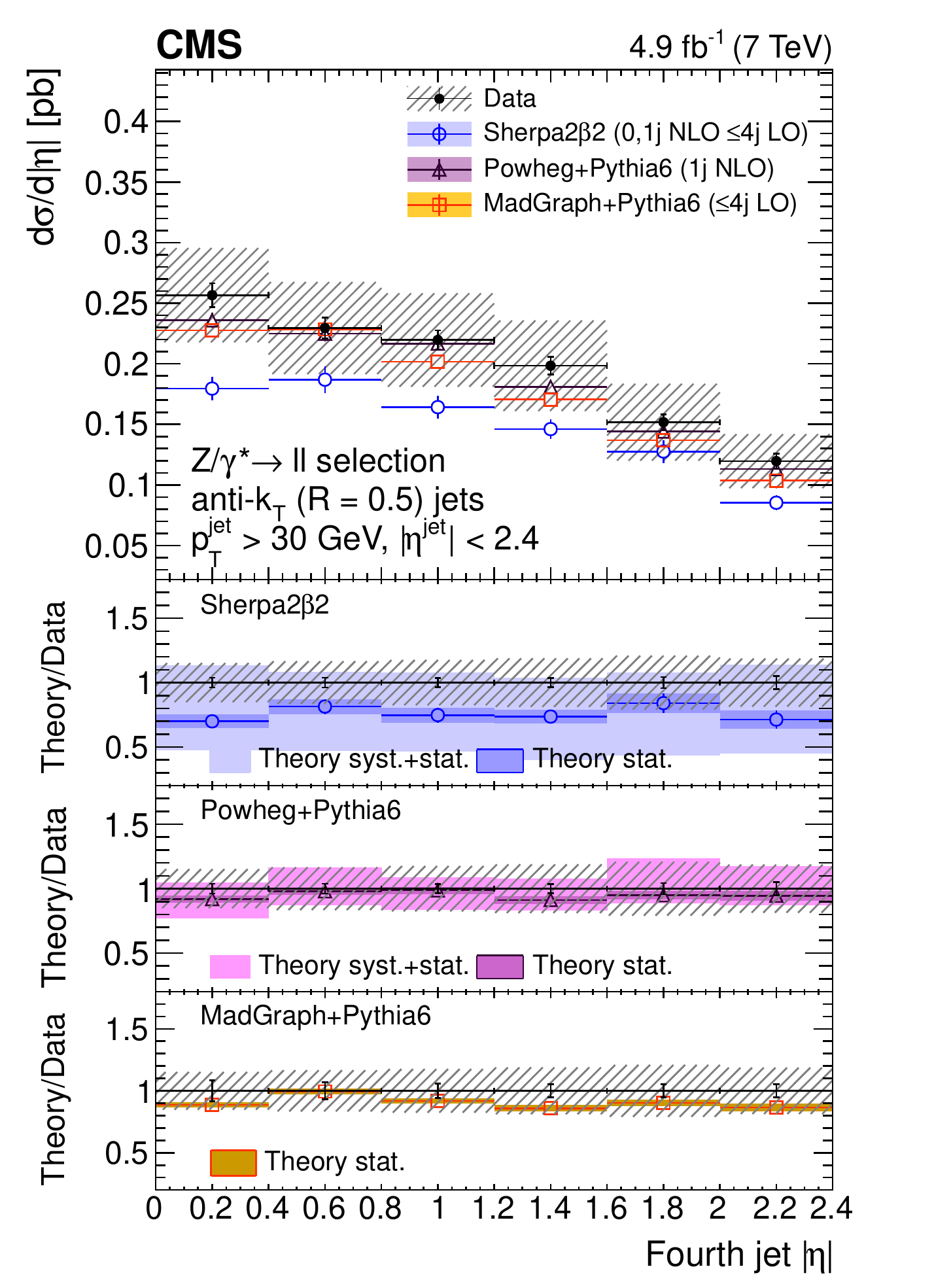}
\caption{Unfolded differential cross section
  as a function of the jet absolute pseudorapidity $\abs{\eta}$
  for the first (top left), second (top right), third (bottom left),
  and fourth (bottom right) highest \pt
  jets, compared with \SHERPA, \POWHEG, and \MADGRAPH
  predictions. Error bars around the experimental points represent the
  statistical uncertainty, while cross-hatched bands represent
  statistical plus systematic uncertainty. The bands around
  theory predictions correspond to the statistical uncertainty of the
  generated sample and, for NLO calculations, to its combination with
  systematic uncertainty related to scale variations.}
\label{FinalJetEta}

\end{figure*}

\begin{figure*}[hbtp]
\centering
\includegraphics[width=\cmsFigWidth]{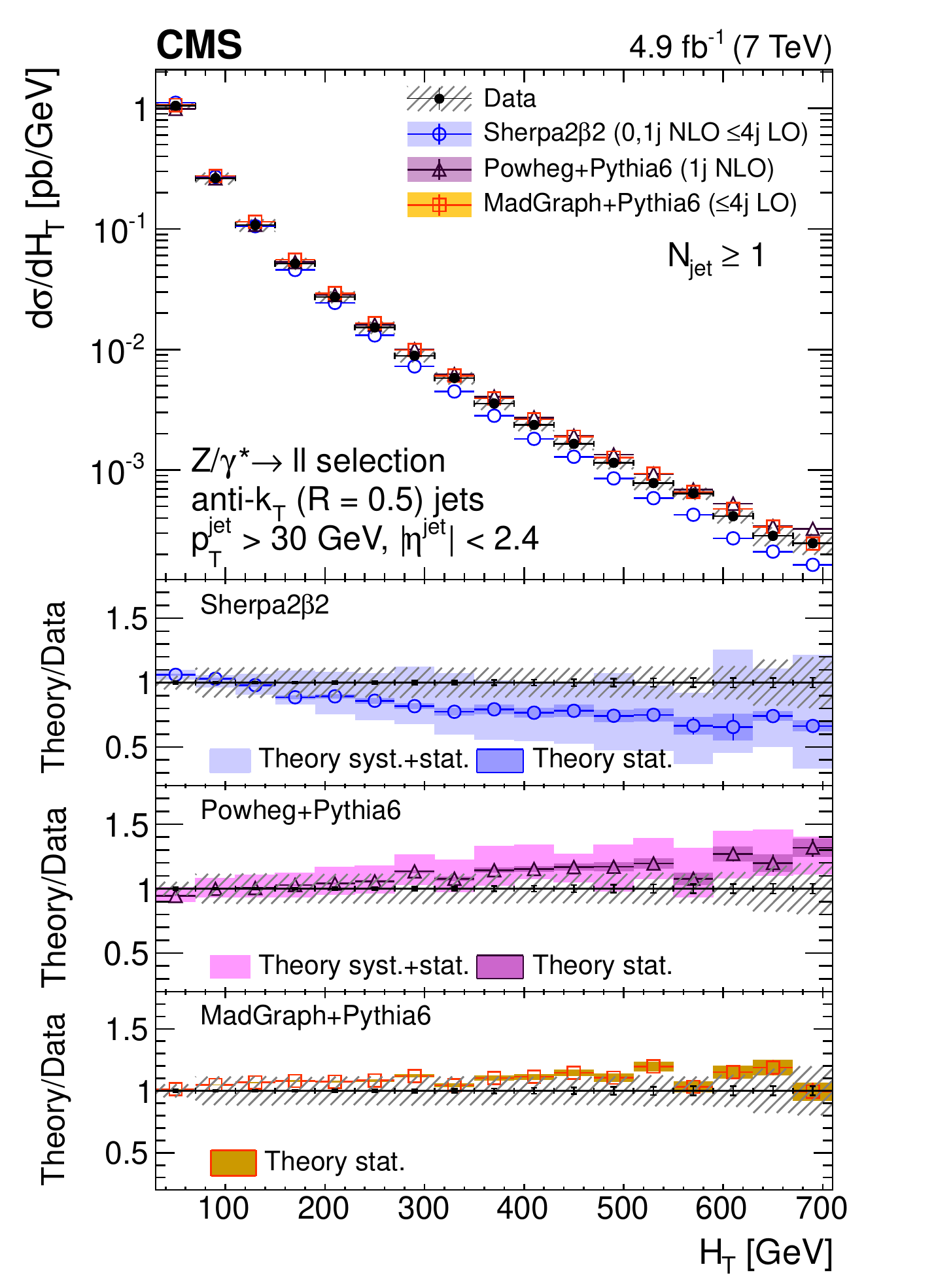}
\includegraphics[width=\cmsFigWidth]{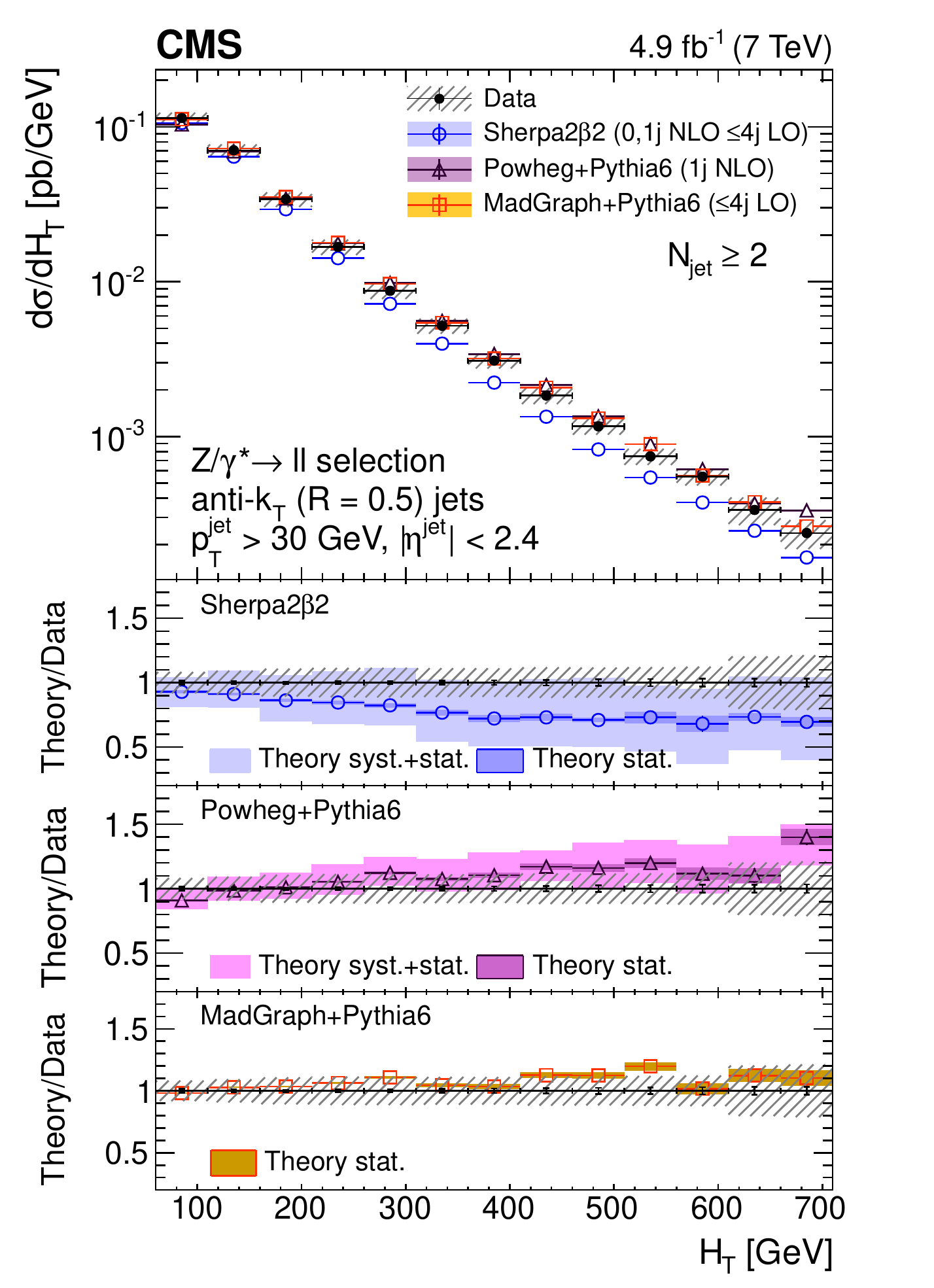}\\
\includegraphics[width=\cmsFigWidth]{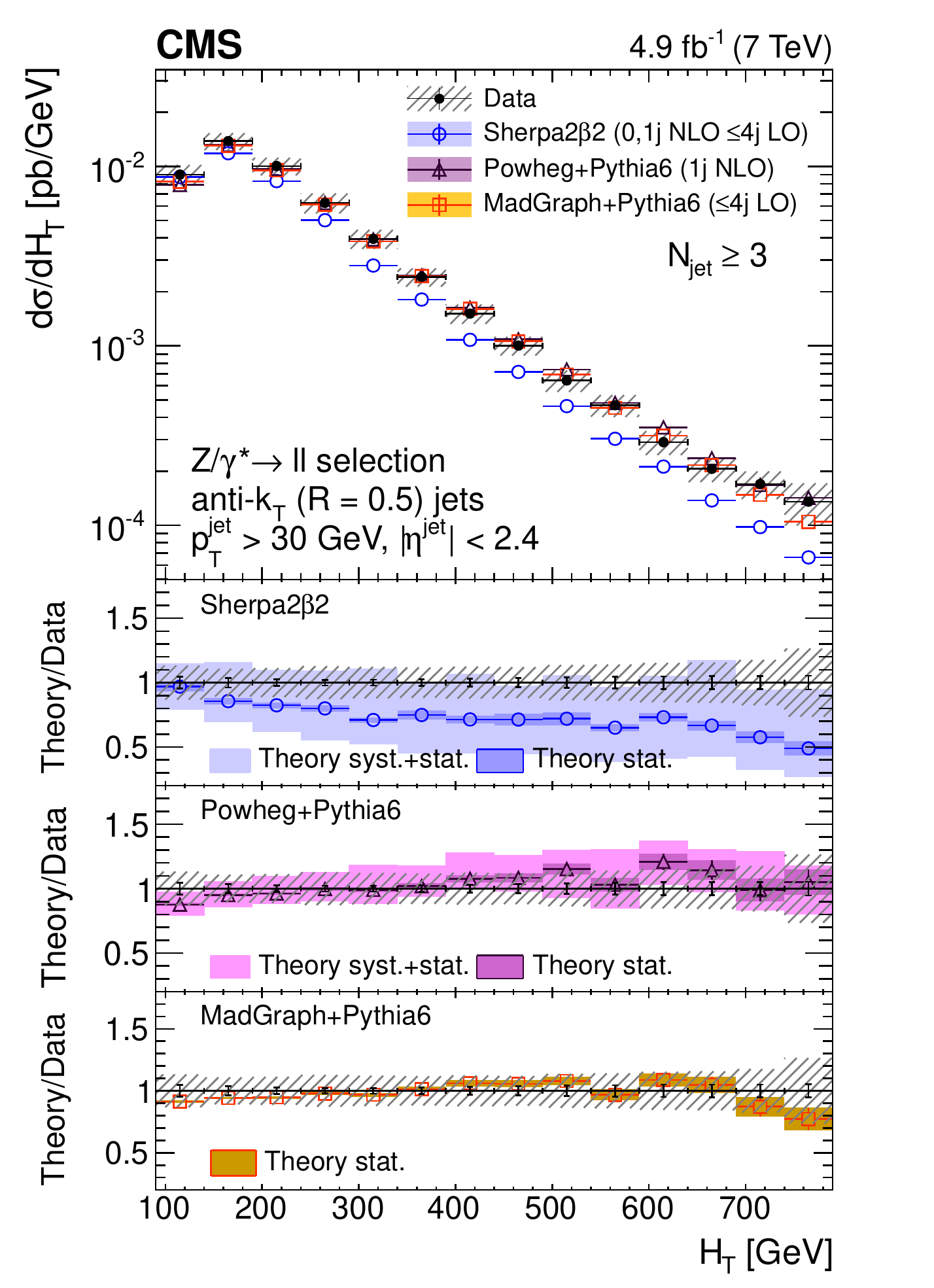}
\includegraphics[width=\cmsFigWidth]{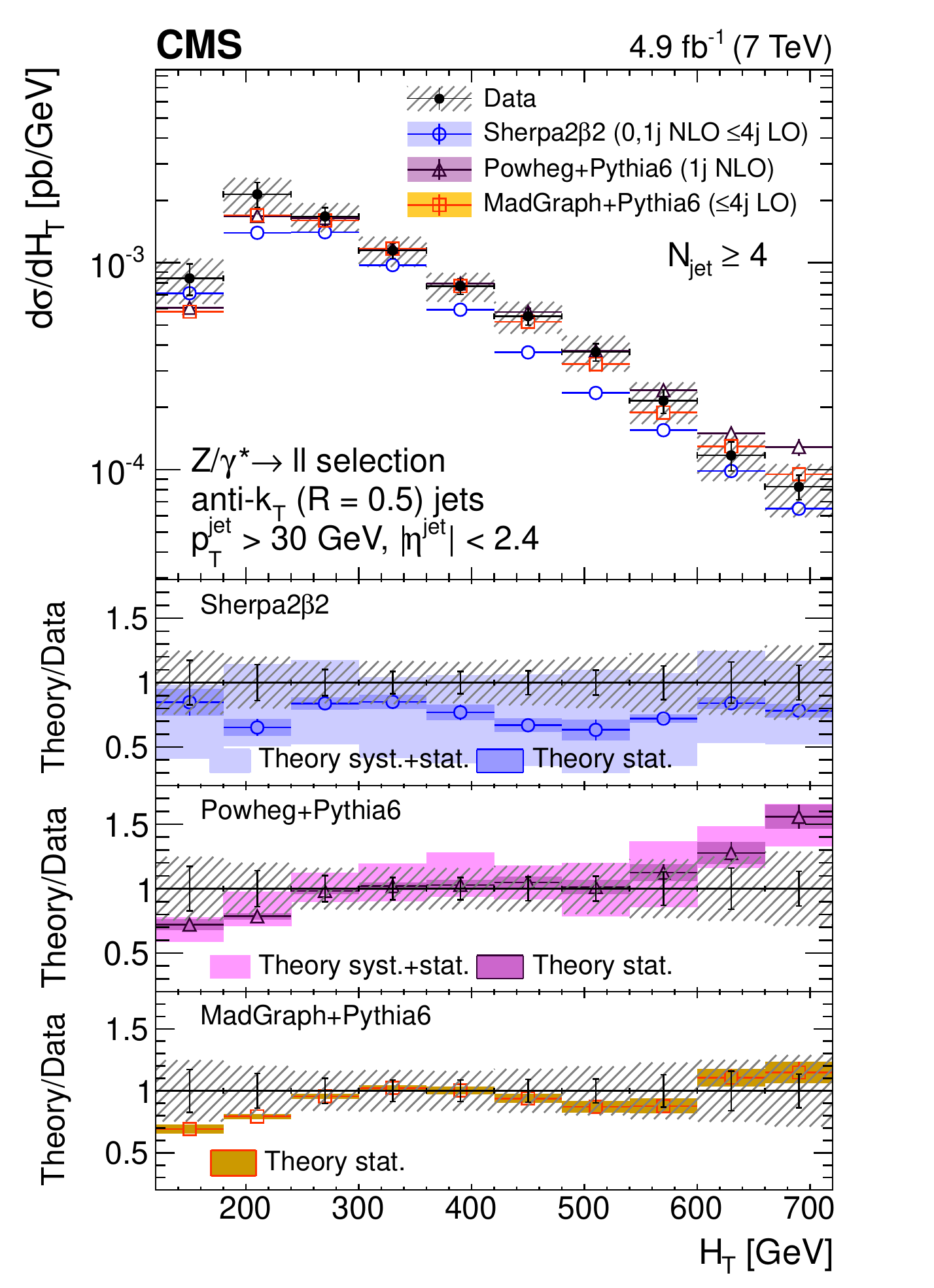}
\caption{Unfolded differential cross section as a function of \HT for
  events with at least one (top left), two (top right), three
  (bottom left), and four (bottom right) jets compared with \SHERPA,
  \POWHEG, and \MADGRAPH predictions. Error bars around the
  experimental points represent the statistical uncertainty, while
  cross-hatched bands represent statistical plus systematic
  uncertainty. The bands around theory predictions correspond to the
  statistical uncertainty of the generated sample and, for NLO
  calculations, to its combination with systematic uncertainty related to
  scale variations.}
\label{FinalJetHT}

\end{figure*}

\begin{figure*}[hbtp]
\centering
\includegraphics[width=\cmsFigWidth]{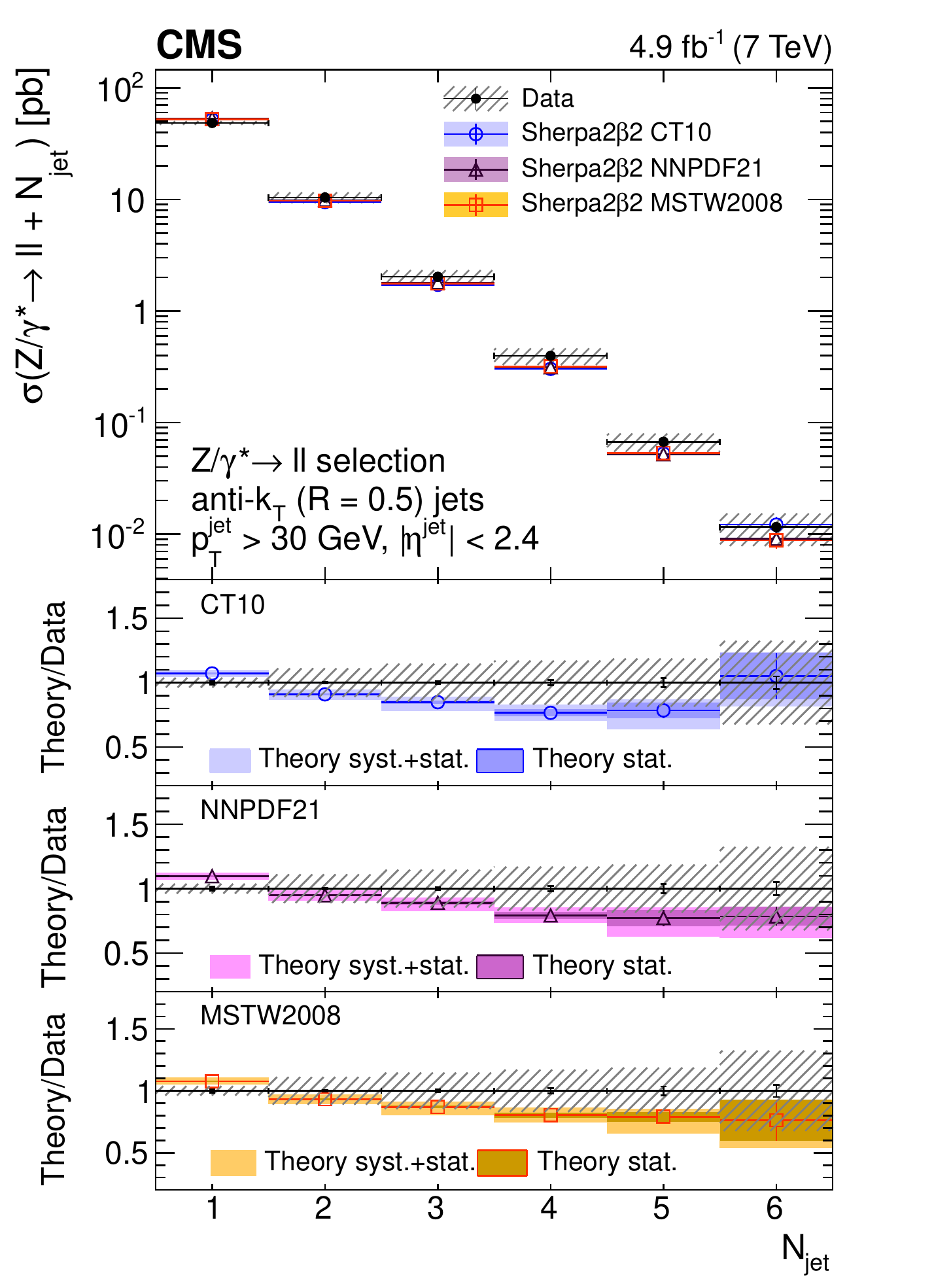}
\includegraphics[width=\cmsFigWidth]{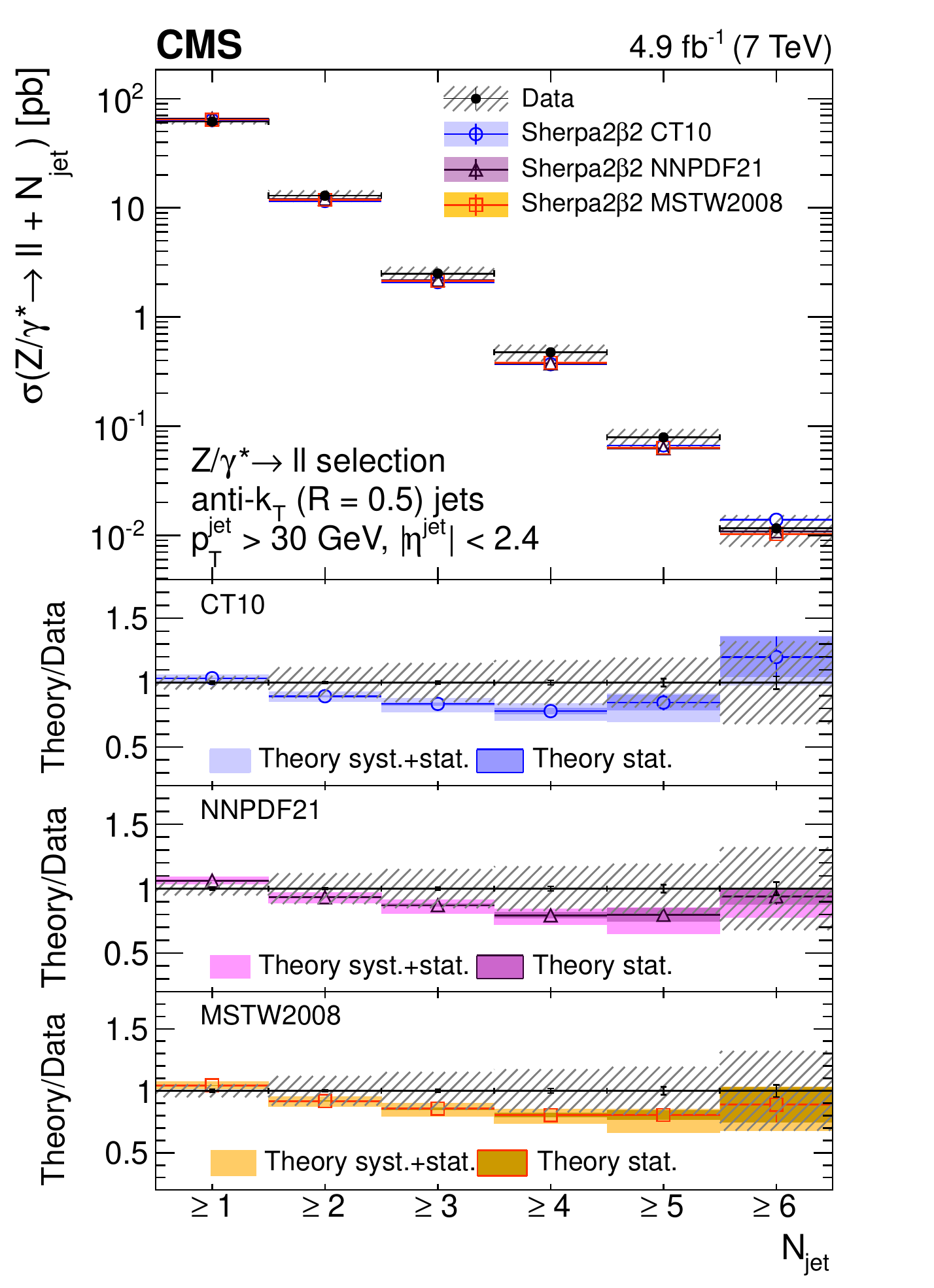}
\caption{Exclusive jet multiplicity distribution~(left) and inclusive jet
  multiplicity distribution~(right), after the unfolding procedure, compared
  with \SHERPA predictions based on the PDF sets CT10, MSTW2008, and NNPDF2.1.
  Error bars around the experimental points represent the statistical
  uncertainty, while cross-hatched bands represent statistical plus
  systematic uncertainty. The bands around theory predictions
  correspond to the statistical uncertainty of the generated sample
  and to its combination with the theoretical PDF uncertainty.}
\label{fig:DiffMultiPDF}

\end{figure*}

\begin{figure*}[hbtp]
\centering
\includegraphics[width=\cmsFigWidth]{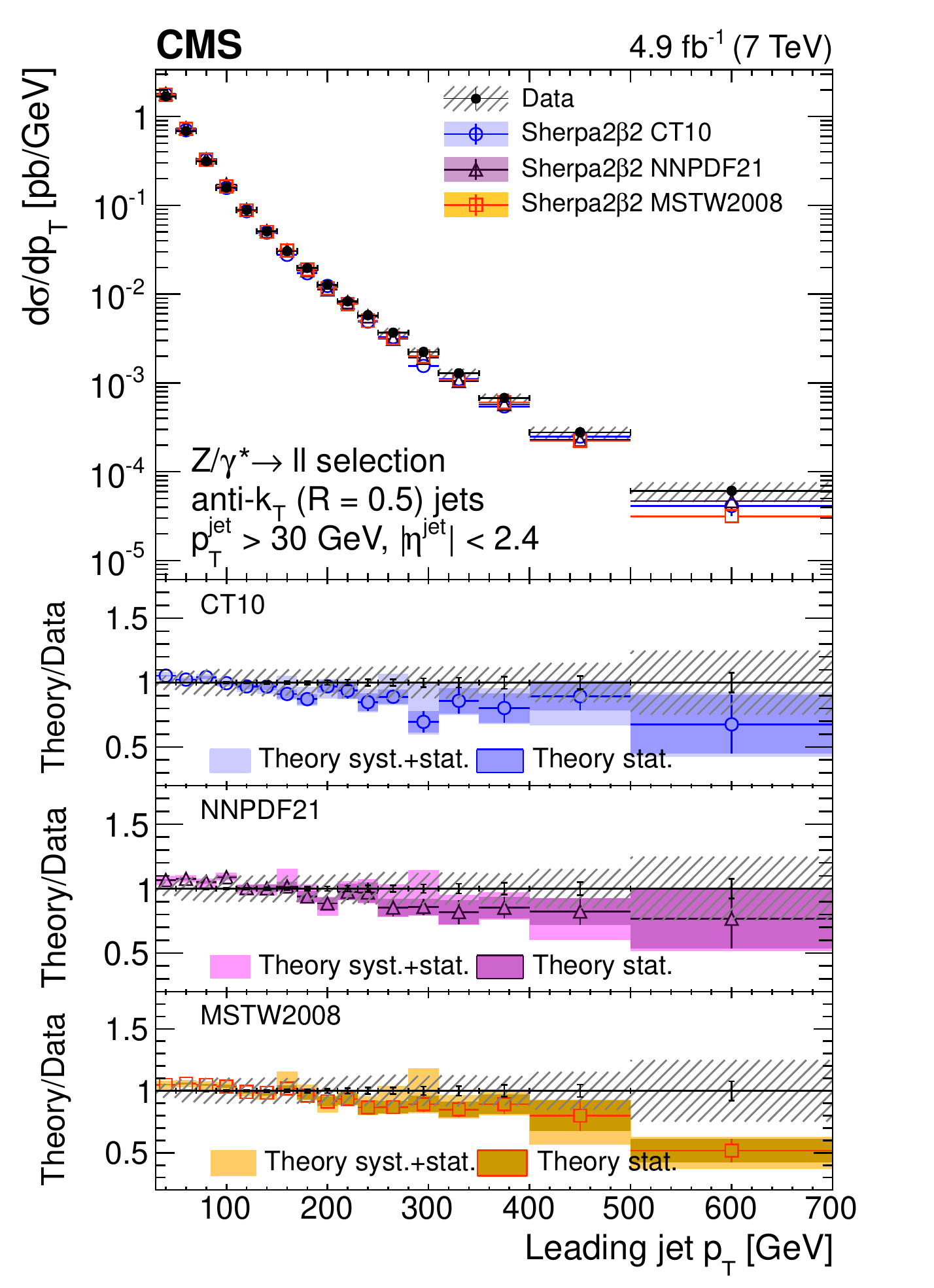}
\includegraphics[width=\cmsFigWidth]{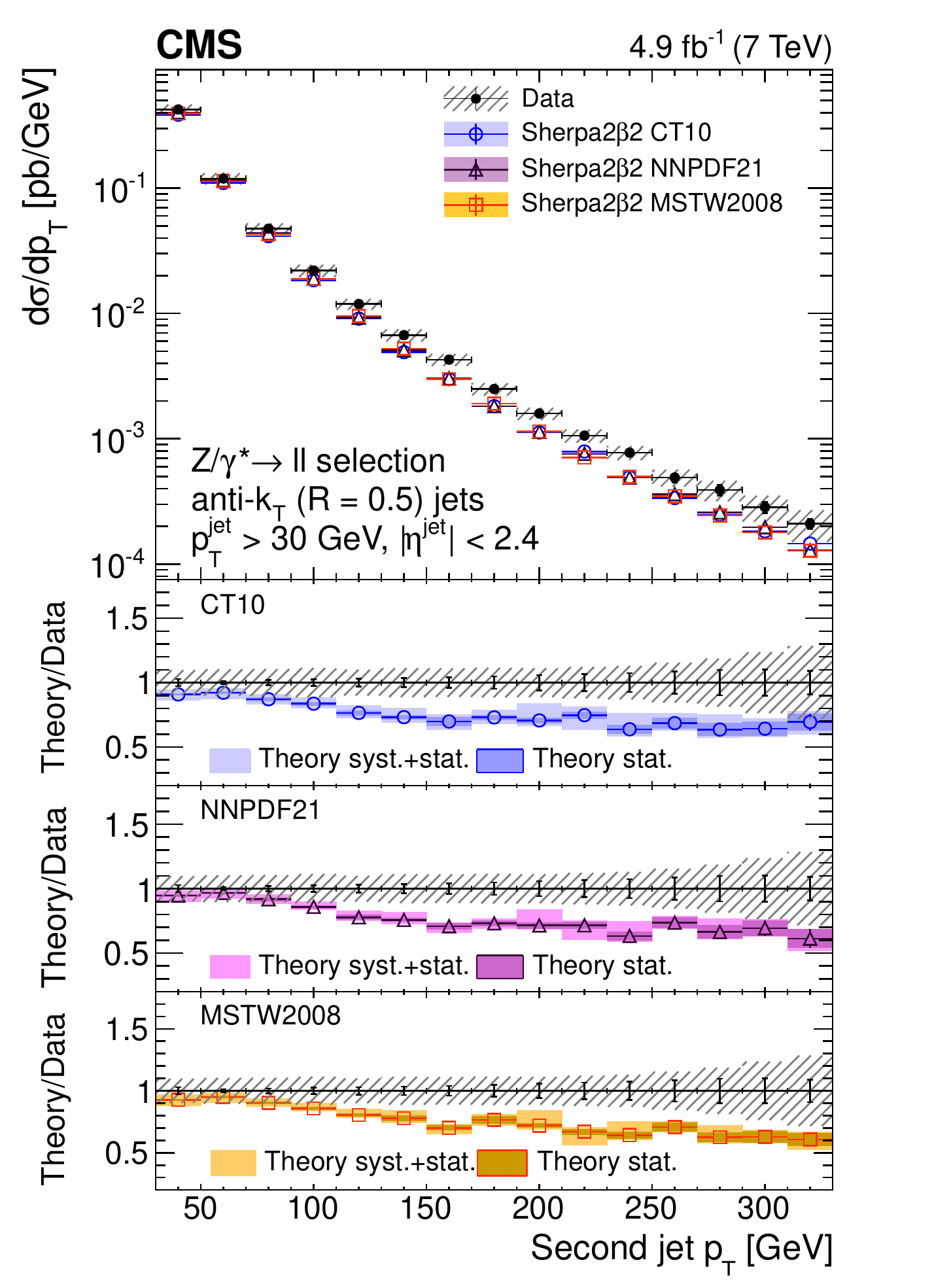}\\
\includegraphics[width=\cmsFigWidth]{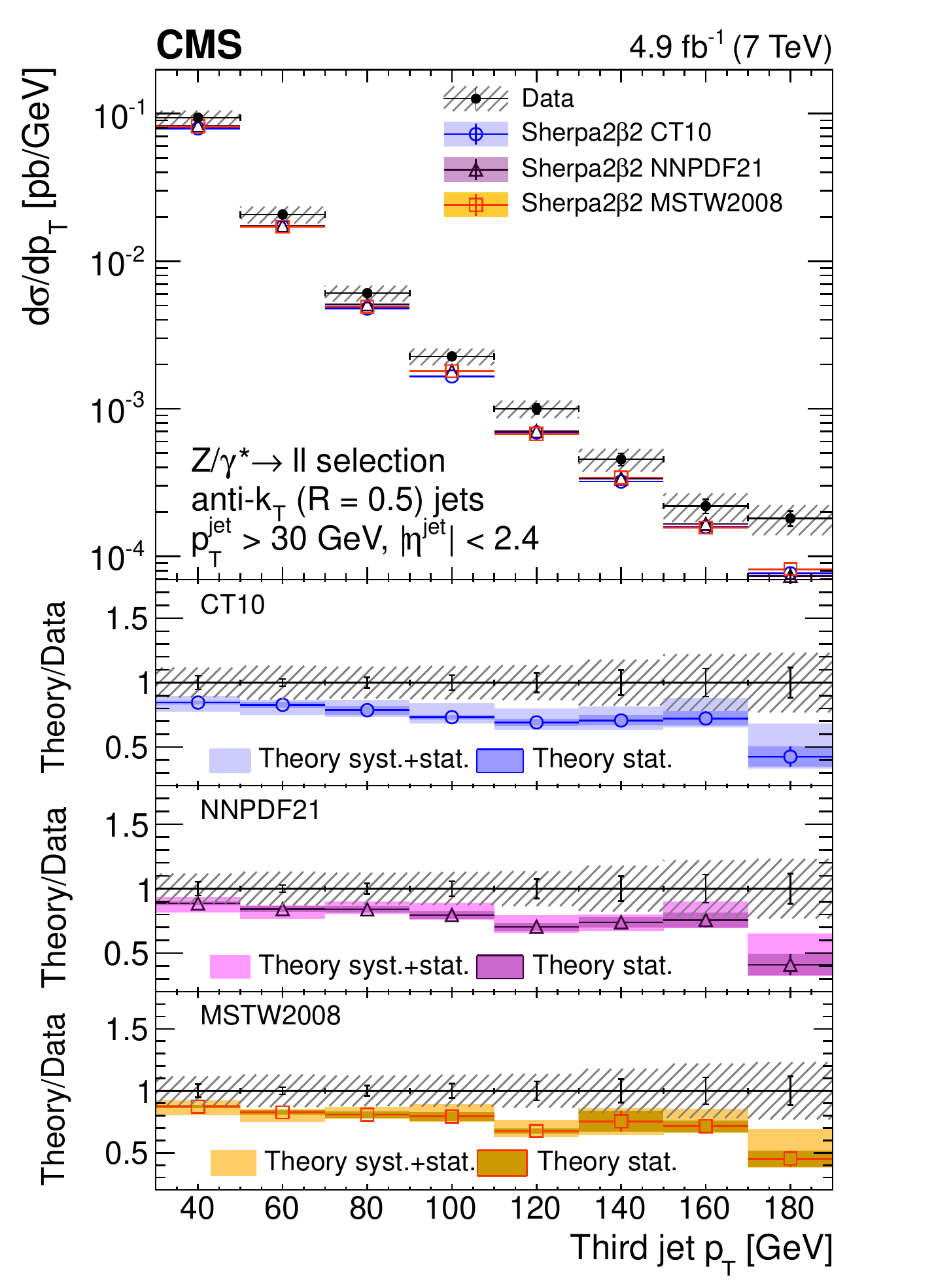}
\includegraphics[width=\cmsFigWidth]{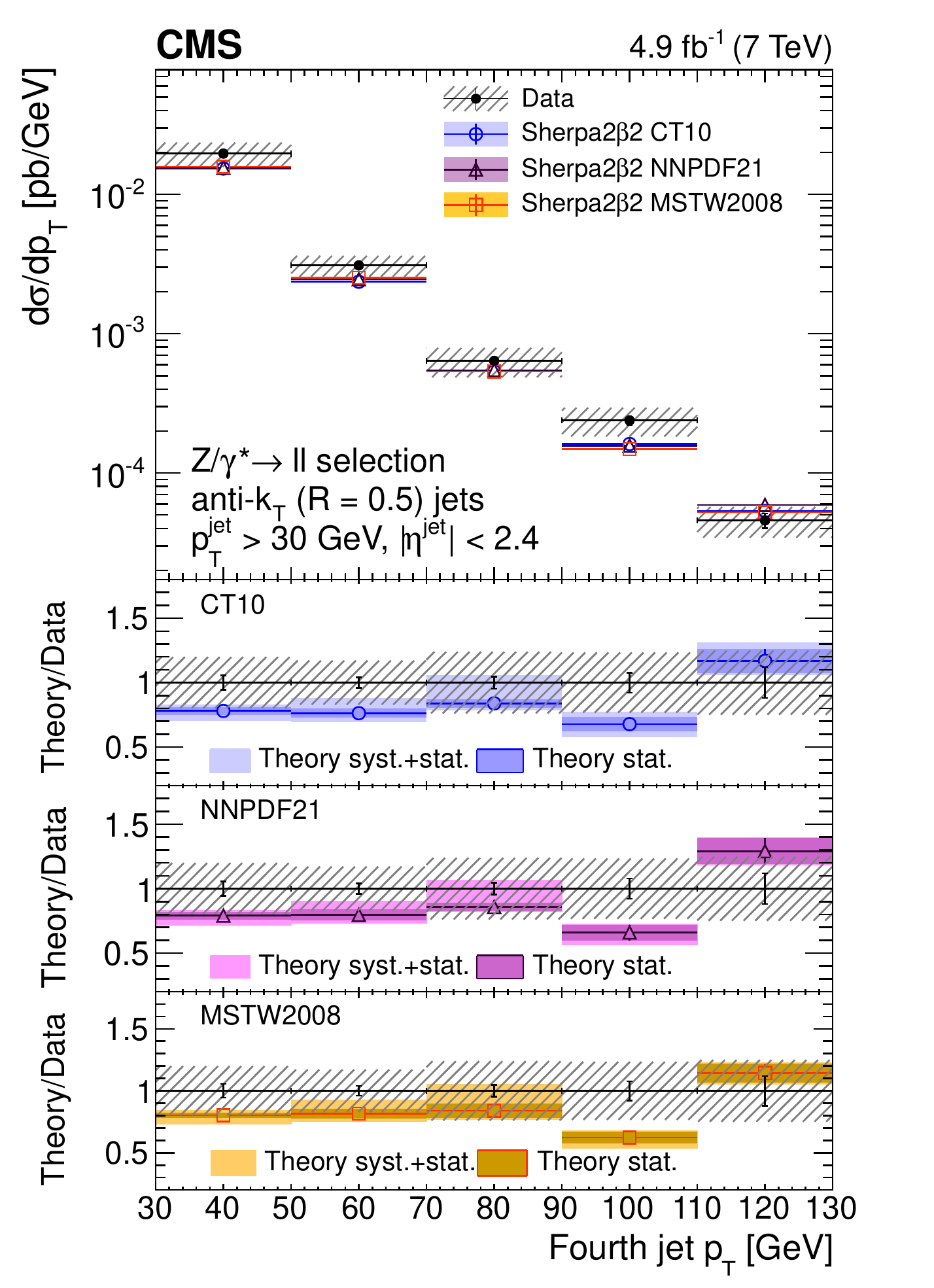}
\caption{Unfolded differential cross section as a function of \pt
  for the first (top left), second (top right), third (bottom left),
  and fourth (bottom right) highest \pt jets, compared with \SHERPA
  predictions based on the PDF sets CT10, MSTW2008, and
  NNPDF2.1. Error bars around the experimental points represent
  the statistical uncertainty, while cross-hatched bands represent
  statistical plus systematic uncertainty. The bands around
  theory predictions correspond to the statistical uncertainty of the
  generated sample and to its combination with the theoretical PDF
  uncertainty.}
\label{FinalJetPtPDF}

\end{figure*}

\begin{figure*}[hbtp]
\centering
\includegraphics[width=\cmsFigWidth]{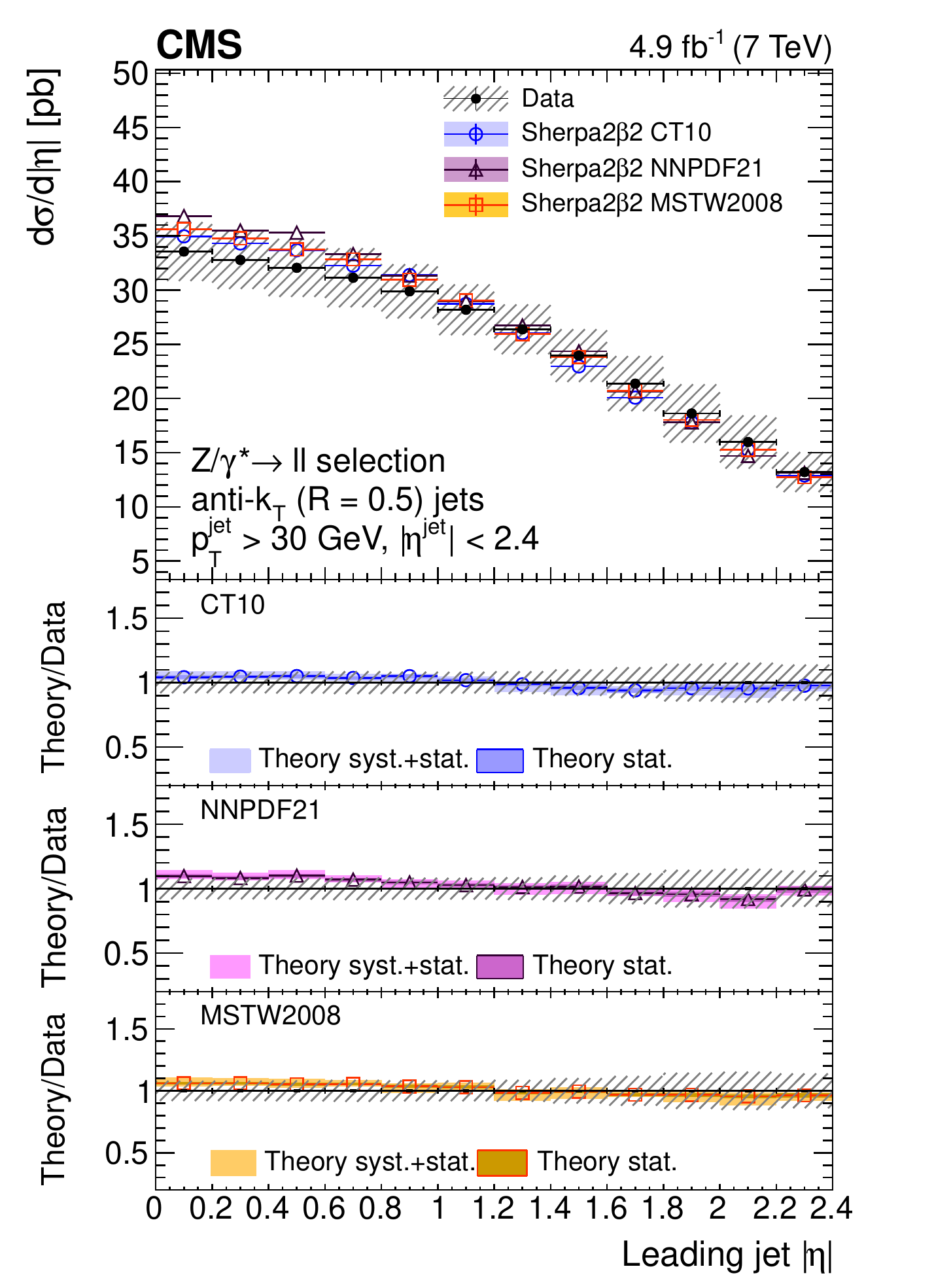}
\includegraphics[width=\cmsFigWidth]{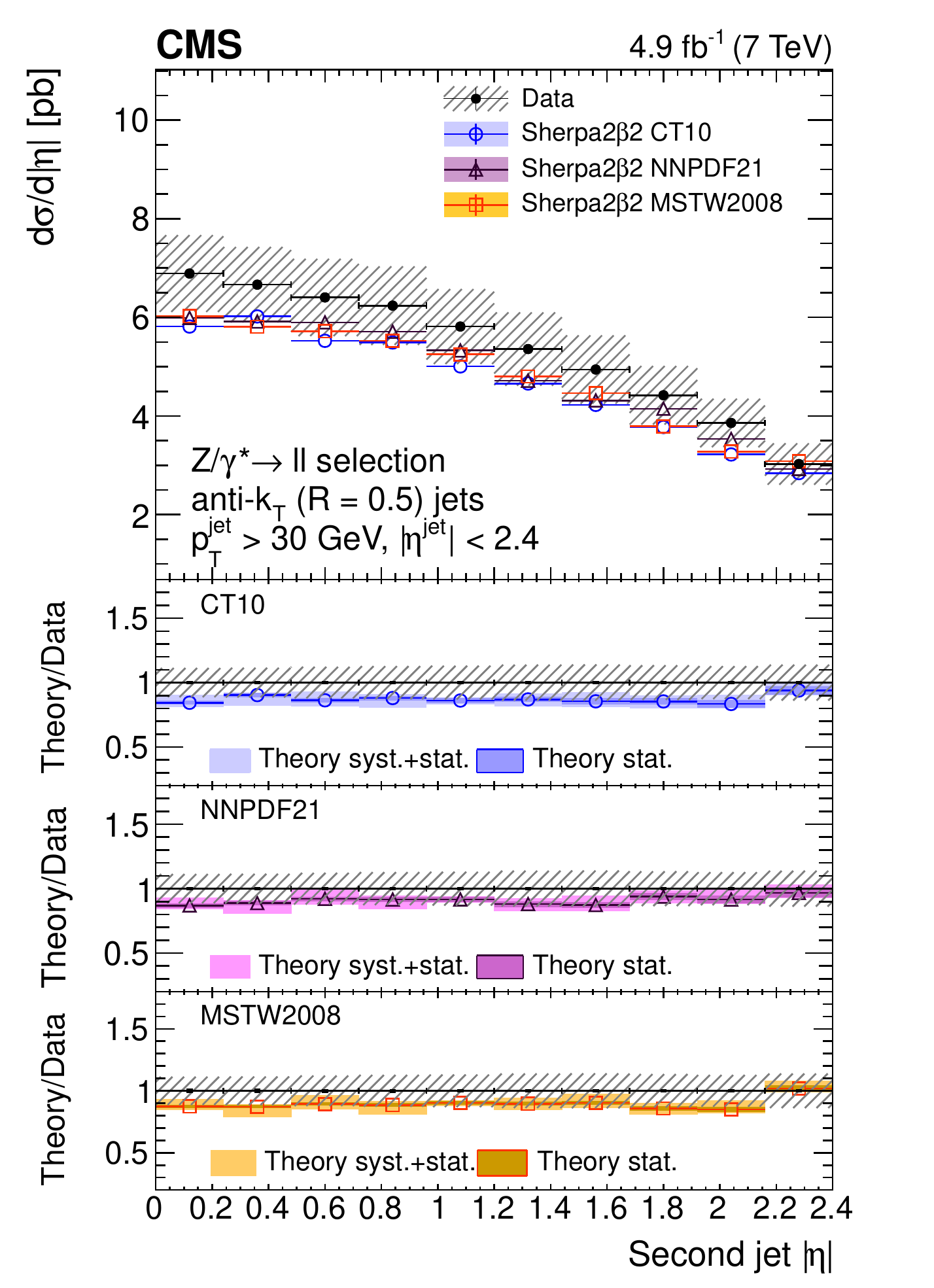}\\
\includegraphics[width=\cmsFigWidth]{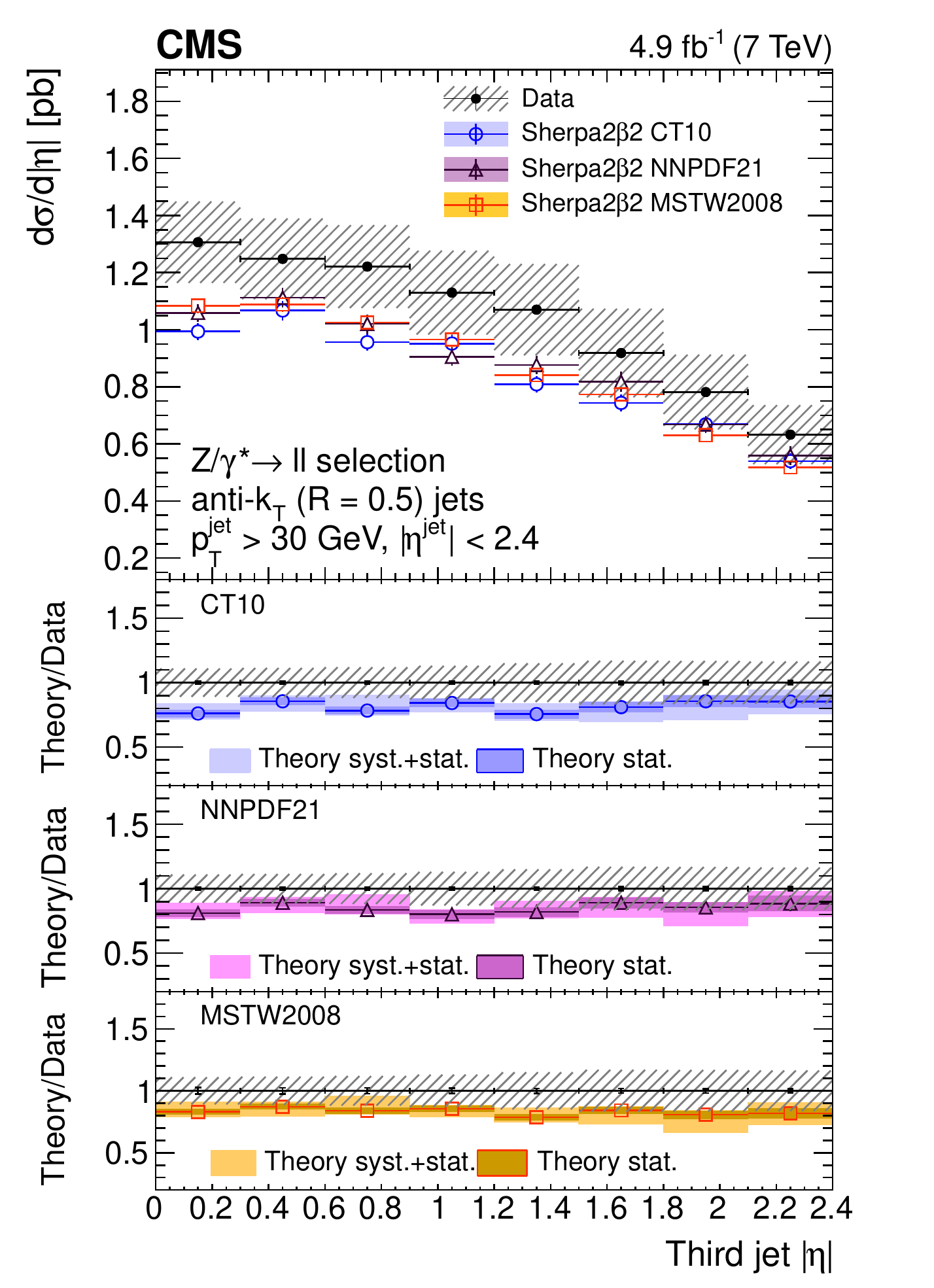}
\includegraphics[width=\cmsFigWidth]{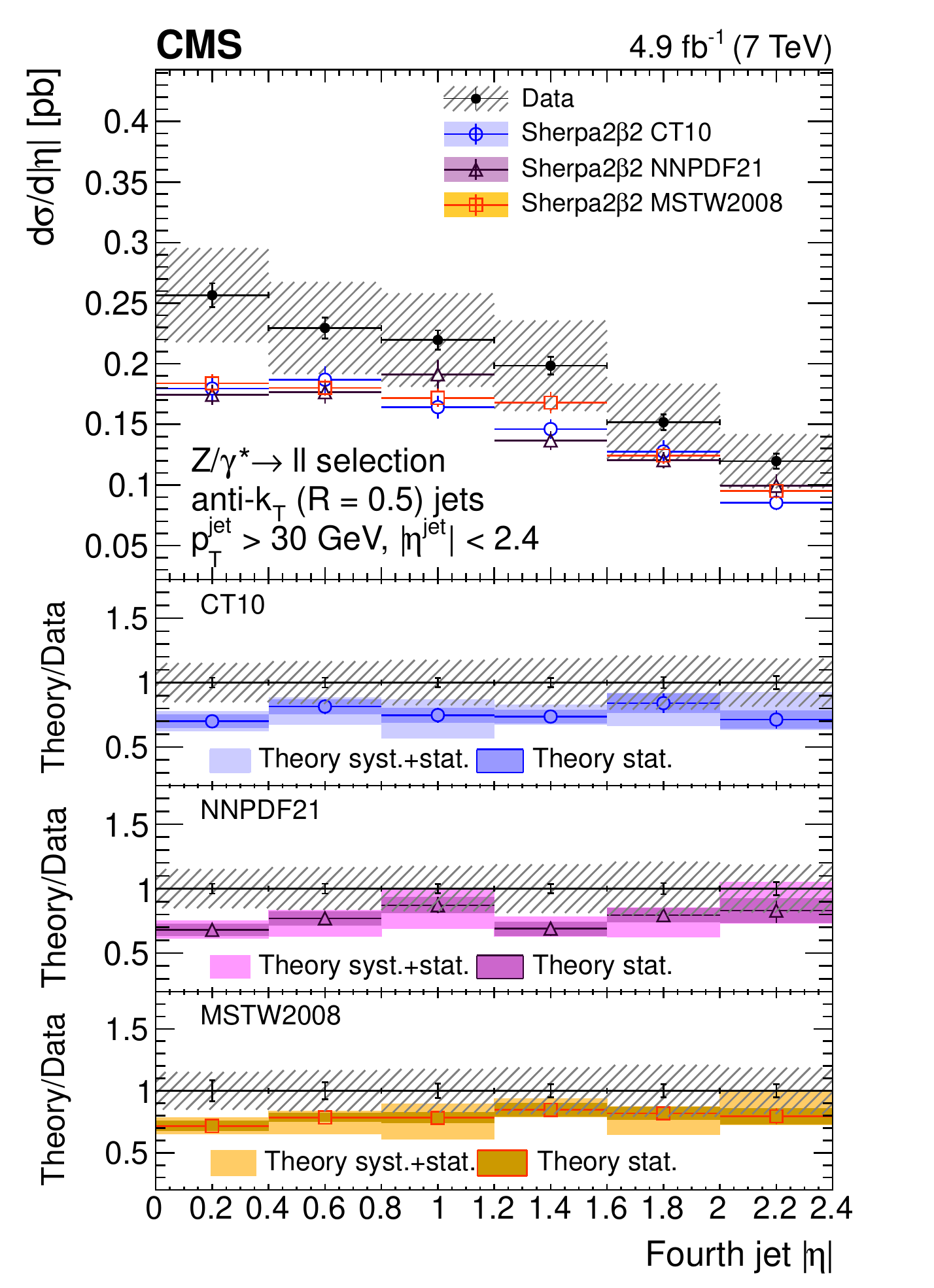}
\caption{Unfolded differential cross section
  as a function of the jet absolute pseudorapidity $\abs{\eta}$
  for the first (top left), second (top right), third (bottom left),
  and fourth (bottom right) highest \pt jets, compared
  with \SHERPA predictions based on the PDF sets CT10, MSTW2008, and NNPDF2.1.
  Error bars around the experimental points represent the statistical
  uncertainty, while cross-hatched bands represent statistical plus
  systematic uncertainty. The bands around theory predictions
  correspond to the statistical uncertainty of the generated sample
  and to its combination with the theoretical PDF uncertainty.}
\label{FinalJetEtaPDF}

\end{figure*}

\begin{figure*}[hbtp]
\centering
\includegraphics[width=\cmsFigWidth]{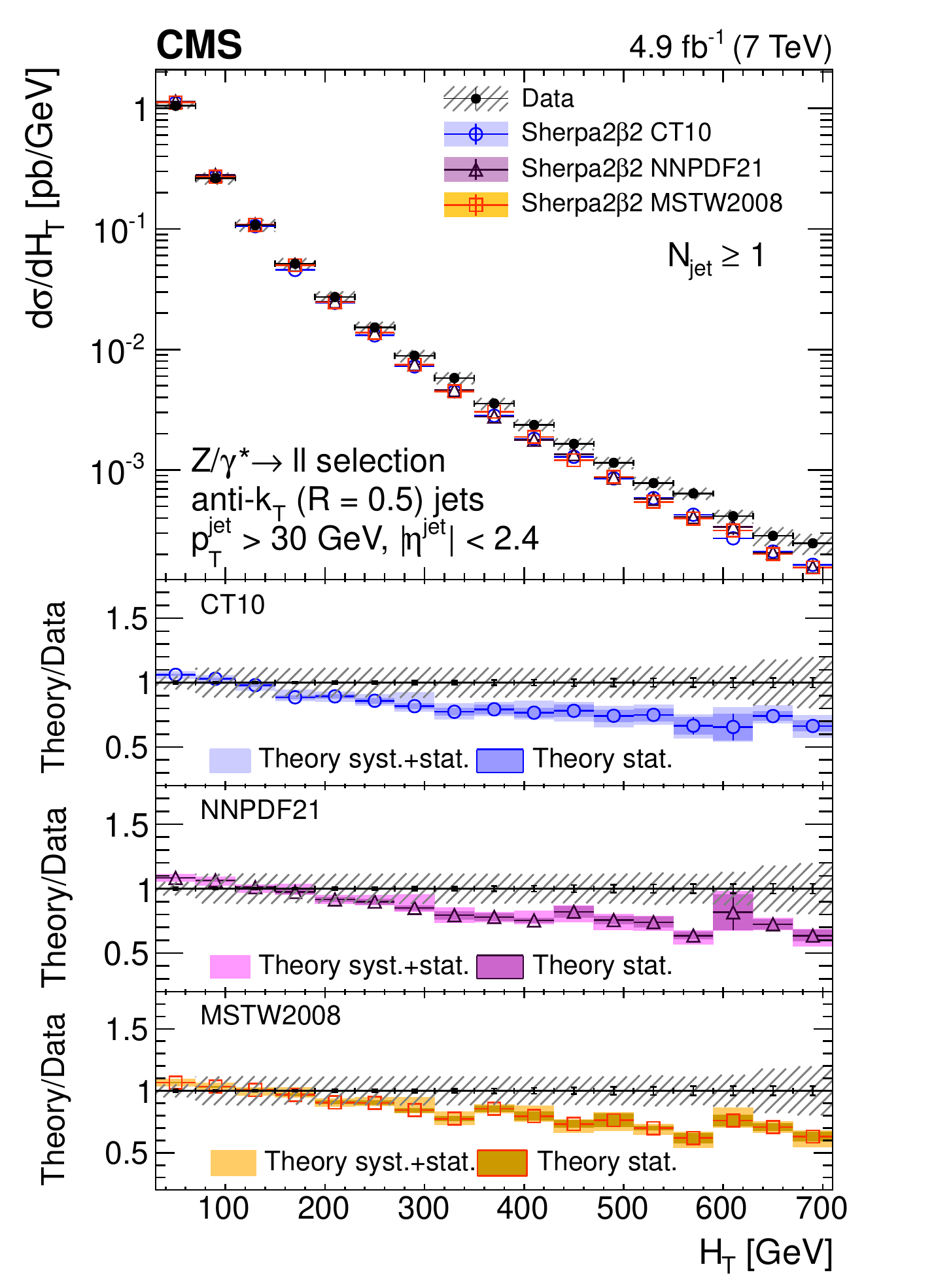}
\includegraphics[width=\cmsFigWidth]{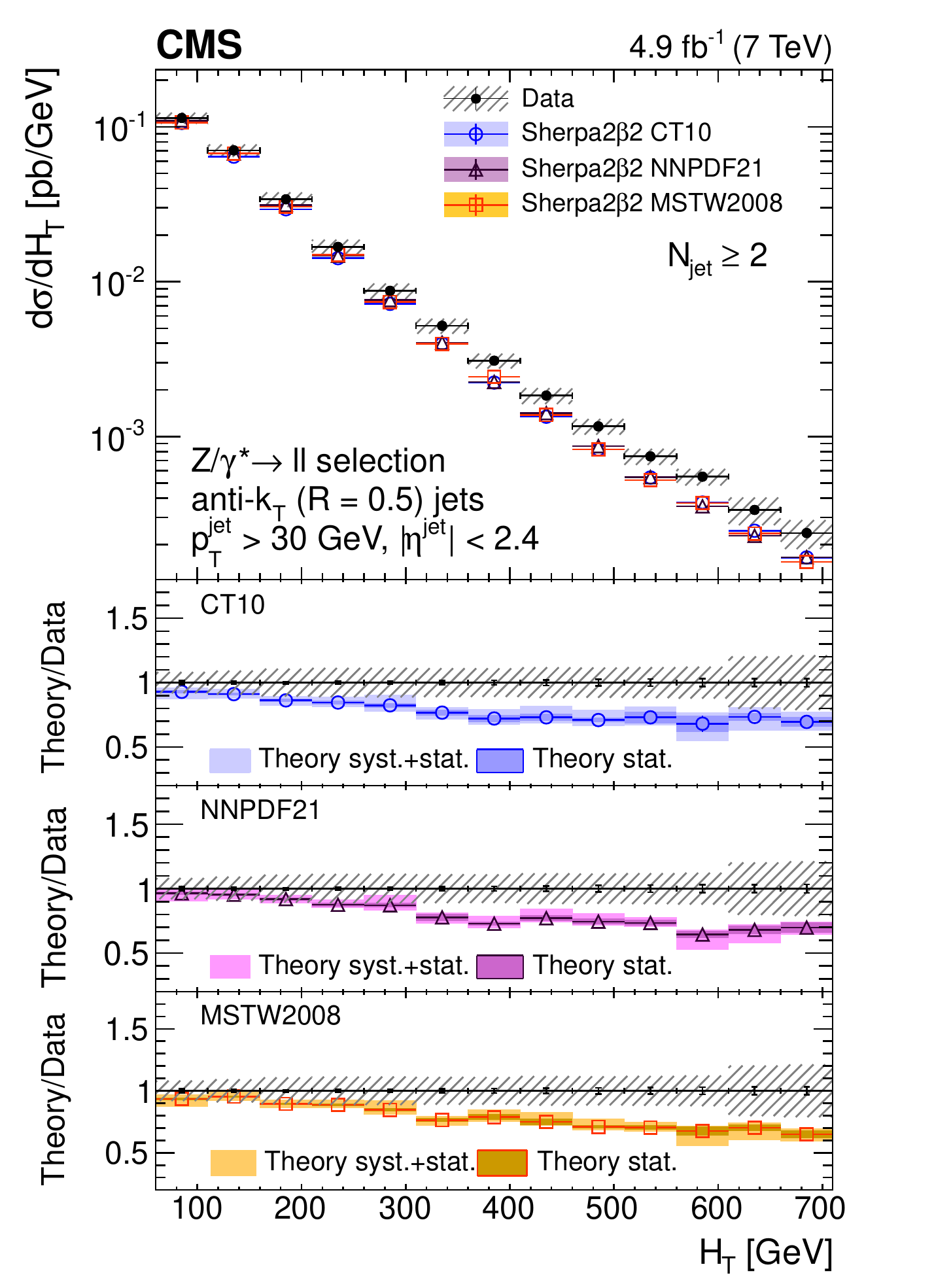}\\
\includegraphics[width=\cmsFigWidth]{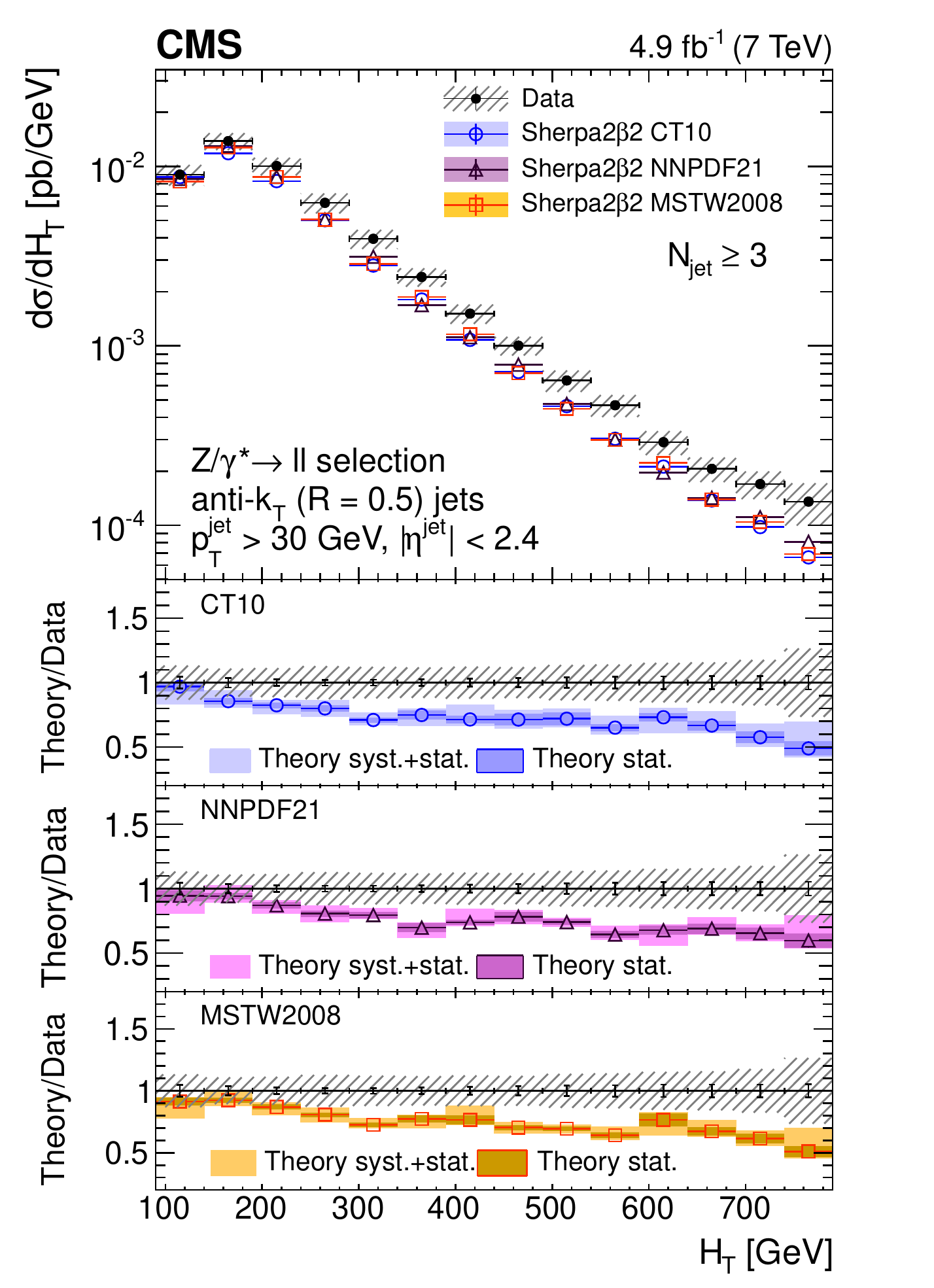}
\includegraphics[width=\cmsFigWidth]{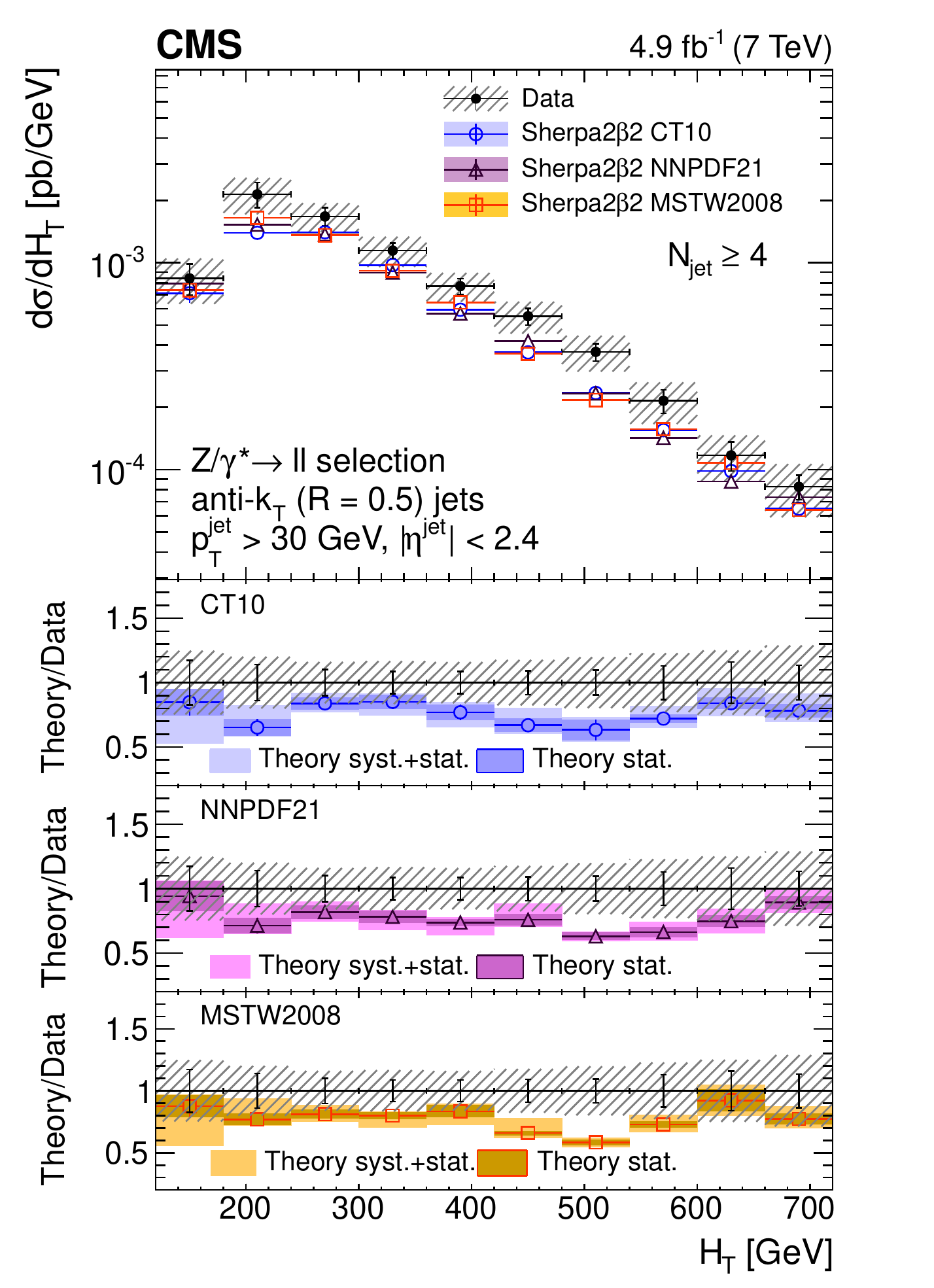}
\caption{Unfolded differential cross section as a function of \HT for
  events with at least one (top left), two (top right), three
  (bottom left), and four (bottom right) jets compared with \SHERPA
  predictions based on the PDF sets CT10, MSTW2008, and
  NNPDF2.1. Error bars around the experimental points represent the
  statistical uncertainty, while cross-hatched bands represent
  statistical plus systematic uncertainty. The bands around theory
  predictions correspond to the statistical uncertainty of the
  generated sample and to its combination with the theoretical
  PDF uncertainty.}
\label{FinalJetHTPDF}

\end{figure*}

The differential cross sections as a function of jet \pt and jet
$\eta$ for the first, second, third, and fourth highest \pt jet in the
event are presented in Figs.~\ref{FinalJetPt} and \ref{FinalJetEta},
respectively. In addition, the differential cross sections as a
function of \HT for events with at least one, two, three, or four jets
are presented in Fig.~\ref{FinalJetHT}.  The pQCD prediction by \MADGRAPH
provides a satisfactory description of data for most
distributions, but shows an excess in the \pt spectra for the first
and second leading jets at $\pt>100$\GeV.  \SHERPA tends to
underestimate the high \pt and \HT regions in most of
the spectra, while remaining compatible with the measurement within
the estimated theoretical uncertainty. \POWHEG predicts harder \pt
spectra than those observed in the data for the events with two or
more jets, where the additional hard radiation is described by the
parton showers and not by matrix elements. This discrepancy is also
reflected in the \HT
distribution. Figures~\ref{fig:DiffMultiPDF}--\ref{FinalJetHTPDF} show
no significant dependence of the level of agreement between
data and the \SHERPA prediction on the PDF set chosen. Hence the PDF
choice cannot explain the observed differences with data.

\section{Summary}

The fiducial production cross section of a \Z boson with at least one
hadronic jet has been measured in proton-proton collisions at
$\sqrt{s}=7$\TeV in a sample corresponding to an integrated luminosity
of 4.9\fbinv.  The measurements comprise inclusive jet multiplicities,
exclusive jet multiplicities, and the differential cross sections as a
function of jet \pt and $\eta$ for the four highest \pt jets of
the event. In addition, the \HT distribution for events with different
minimum numbers of jets has been measured. All measured differential
cross sections are corrected for detector effects and compared with
theoretical predictions at particle level.

The predictions of calculations combining matrix element and parton
shower can describe, within uncertainties, the measured spectra over
a wide kinematical range.  The measured jet multiplicity distributions
and their NLO theoretical predictions from the \SHERPA and \POWHEG
generators are consistent within the experimental and theoretical
uncertainties. However, \SHERPA predicts softer \pt and \HT spectra
than the measured ones, while \POWHEG shows an excess compared to data
in the high \pt and \HT regions.  In particular, the \POWHEG spectra
are harder for the highest jet multiplicities, which are described
only by parton showers.  The tree level calculation based on \MADGRAPH
predicts harder \pt spectra than the measured ones for low jet
multiplicities. 

\begin{acknowledgments}
\hyphenation{Bundes-ministerium Forschungs-gemeinschaft Forschungs-zentren} We congratulate our colleagues in the CERN accelerator departments for the excellent performance of the LHC and thank the technical and administrative staffs at CERN and at other CMS institutes for their contributions to the success of the CMS effort. In addition, we gratefully acknowledge the computing centres and personnel of the Worldwide LHC Computing Grid for delivering so effectively the computing infrastructure essential to our analyses. Finally, we acknowledge the enduring support for the construction and operation of the LHC and the CMS detector provided by the following funding agencies: the Austrian Federal Ministry of Science, Research and Economy and the Austrian Science Fund; the Belgian Fonds de la Recherche Scientifique, and Fonds voor Wetenschappelijk Onderzoek; the Brazilian Funding Agencies (CNPq, CAPES, FAPERJ, and FAPESP); the Bulgarian Ministry of Education and Science; CERN; the Chinese Academy of Sciences, Ministry of Science and Technology, and National Natural Science Foundation of China; the Colombian Funding Agency (COLCIENCIAS); the Croatian Ministry of Science, Education and Sport, and the Croatian Science Foundation; the Research Promotion Foundation, Cyprus; the Ministry of Education and Research, Estonian Research Council via IUT23-4 and IUT23-6 and European Regional Development Fund, Estonia; the Academy of Finland, Finnish Ministry of Education and Culture, and Helsinki Institute of Physics; the Institut National de Physique Nucl\'eaire et de Physique des Particules~/~CNRS, and Commissariat \`a l'\'Energie Atomique et aux \'Energies Alternatives~/~CEA, France; the Bundesministerium f\"ur Bildung und Forschung, Deutsche Forschungsgemeinschaft, and Helmholtz-Gemeinschaft Deutscher Forschungszentren, Germany; the General Secretariat for Research and Technology, Greece; the National Scientific Research Foundation, and National Innovation Office, Hungary; the Department of Atomic Energy and the Department of Science and Technology, India; the Institute for Studies in Theoretical Physics and Mathematics, Iran; the Science Foundation, Ireland; the Istituto Nazionale di Fisica Nucleare, Italy; the Korean Ministry of Education, Science and Technology and the World Class University program of NRF, Republic of Korea; the Lithuanian Academy of Sciences; the Ministry of Education, and University of Malaya (Malaysia); the Mexican Funding Agencies (CINVESTAV, CONACYT, SEP, and UASLP-FAI); the Ministry of Business, Innovation and Employment, New Zealand; the Pakistan Atomic Energy Commission; the Ministry of Science and Higher Education and the National Science Centre, Poland; the Funda\c{c}\~ao para a Ci\^encia e a Tecnologia, Portugal; JINR, Dubna; the Ministry of Education and Science of the Russian Federation, the Federal Agency of Atomic Energy of the Russian Federation, Russian Academy of Sciences, and the Russian Foundation for Basic Research; the Ministry of Education, Science and Technological Development of Serbia; the Secretar\'{\i}a de Estado de Investigaci\'on, Desarrollo e Innovaci\'on and Programa Consolider-Ingenio 2010, Spain; the Swiss Funding Agencies (ETH Board, ETH Zurich, PSI, SNF, UniZH, Canton Zurich, and SER); the Ministry of Science and Technology, Taipei; the Thailand Center of Excellence in Physics, the Institute for the Promotion of Teaching Science and Technology of Thailand, Special Task Force for Activating Research and the National Science and Technology Development Agency of Thailand; the Scientific and Technical Research Council of Turkey, and Turkish Atomic Energy Authority; the National Academy of Sciences of Ukraine, and State Fund for Fundamental Researches, Ukraine; the Science and Technology Facilities Council, UK; the US Department of Energy, and the US National Science Foundation.

Individuals have received support from the Marie-Curie programme and the European Research Council and EPLANET (European Union); the Leventis Foundation; the A. P. Sloan Foundation; the Alexander von Humboldt Foundation; the Belgian Federal Science Policy Office; the Fonds pour la Formation \`a la Recherche dans l'Industrie et dans l'Agriculture (FRIA-Belgium); the Agentschap voor Innovatie door Wetenschap en Technologie (IWT-Belgium); the Ministry of Education, Youth and Sports (MEYS) of the Czech Republic; the Council of Science and Industrial Research, India; the HOMING PLUS programme of Foundation for Polish Science, cofinanced from European Union, Regional Development Fund; the Compagnia di San Paolo (Torino); the Consorzio per la Fisica (Trieste); MIUR project 20108T4XTM (Italy); the Thalis and Aristeia programmes cofinanced by EU-ESF and the Greek NSRF; and the National Priorities Research Program by Qatar National Research Fund.
\end{acknowledgments}

\bibliography{auto_generated}   

\providecommand{\href}[2]{#2}\begingroup\raggedright\begin{thebibliography}{10}%
\makeatletter
\providecommand{\hrefCMSnoop }[0]{\@secondoftwo}%
\makeatother
\providecommand{\doi}{\texttt{doi:}\begingroup \urlstyle{tt}\Url}

\bibitem{Alwall:2011uj}
J.~Alwall\hrefCMSnoop {}{ {et~al.}, ``{MadGraph} 5: going beyond'',} \textit{
  JHEP} \textbf{ 06} (2011) 128,
  \href{http://dx.doi.org/10.1007/JHEP06(2011)128}{\doi{10.1007/JHEP06(2011)128}},
\href{http://www.arXiv.org/abs/1106.0522}{\texttt{ arXiv:1106.0522}}.

\bibitem{Nason:2004rx}
\hrefCMSnoop {}{P.~Nason, ``A new method for combining {NLO QCD} with shower
  {Monte Carlo} algorithms'',} \textit{ JHEP} \textbf{ 11} (2004) 040,
  \href{http://dx.doi.org/10.1088/1126-6708/2004/11/040}{\doi{10.1088/1126-6708/2004/11/040}},
\href{http://www.arXiv.org/abs/hep-ph/0409146}{\texttt{ arXiv:hep-ph/0409146}}.

\bibitem{Frixione:2007vw}
\hrefCMSnoop {}{S.~Frixione, P.~Nason, and C.~Oleari, ``Matching {NLO QCD}
  computations with parton shower simulations: the {POWHEG} method'',} \textit{
  JHEP} \textbf{ 11} (2007) 070,
  \href{http://dx.doi.org/10.1088/1126-6708/2007/11/070}{\doi{10.1088/1126-6708/2007/11/070}},
\href{http://www.arXiv.org/abs/0709.2092}{\texttt{ arXiv:0709.2092}}.

\bibitem{Alioli:2010xd}
\hrefCMSnoop {}{S.~Alioli, P.~Nason, C.~Oleari, and E.~Re, ``A general
  framework for implementing {NLO} calculations in shower {Monte Carlo}
  programs: the {POWHEG BOX}'',} \textit{ JHEP} \textbf{ 06} (2010) 043,
  \href{http://dx.doi.org/10.1007/JHEP06(2010)043}{\doi{10.1007/JHEP06(2010)043}},
\href{http://www.arXiv.org/abs/1002.2581}{\texttt{ arXiv:1002.2581}}.

\bibitem{Chatrchyan:2013zna}
\hrefCMSnoop {}{{CMS} Collaboration, ``Search for the standard model {Higgs}
  boson produced in association with a {W} or a {Z} boson and decaying to
  bottom quarks'',} \textit{ Phys. Rev. D} \textbf{ 89} (2014) 012003,
  \href{http://dx.doi.org/10.1103/PhysRevD.89.012003}{\doi{10.1103/PhysRevD.89.012003}},
  \href{http://www.arXiv.org/abs/1310.3687}{\texttt{ arXiv:1310.3687}}.

\bibitem{Aad:2014xzb}
\hrefCMSnoop {}{{ATLAS} Collaboration, ``Search for the \bbbar decay of the
  {Standard Model Higgs} boson in associated (W/Z)H production with the {ATLAS}
  detector'',} \textit{ JHEP} \textbf{ 01} (2015) 069,
  \href{http://dx.doi.org/10.1007/JHEP01(2015)069}{\doi{10.1007/JHEP01(2015)069}},
\href{http://www.arXiv.org/abs/1409.6212}{\texttt{ arXiv:1409.6212}}.

\bibitem{Chatrchyan:2013tna}
\hrefCMSnoop {}{{CMS} Collaboration, ``Event shapes and azimuthal correlations
  in Z + jets events in $\Pp\Pp$ collisions at $\sqrt{s}=7$ TeV'',} \textit{
  Phys. Lett. B} \textbf{ 722} (2013) 238,
  \href{http://dx.doi.org/10.1016/j.physletb.2013.04.025}{\doi{10.1016/j.physletb.2013.04.025}},
\href{http://www.arXiv.org/abs/1301.1646}{\texttt{ arXiv:1301.1646}}.

\bibitem{Aaltonen:2007ae}
\hrefCMSnoop {}{{CDF} Collaboration, ``Measurement of inclusive jet
  cross-sections in Z/$\Pgg^*(\rightarrow \Pep \Pem$) + jets production in $\Pp
  \bar{\Pp}$ collisions at $\sqrt{s}$ = 1.96 TeV'',} \textit{ Phys. Rev. Lett.}
  \textbf{ 100} (2008) 102001,
  \href{http://dx.doi.org/10.1103/PhysRevLett.100.102001}{\doi{10.1103/PhysRevLett.100.102001}},
\href{http://www.arXiv.org/abs/0711.3717}{\texttt{ arXiv:0711.3717}}.

\bibitem{Abazov:2009av}
\hrefCMSnoop {}{{D0} Collaboration, ``Measurements of differential cross
  sections of Z/$\Pgg^*$+jets+X events in $\Pp \bar{\Pp}$ collisions at
  $\sqrt{s}$ = 1.96 TeV'',} \textit{ Phys. Lett. B} \textbf{ 678} (2009) 45,
  \href{http://dx.doi.org/10.1016/j.physletb.2009.05.058}{\doi{10.1016/j.physletb.2009.05.058}},
\href{http://www.arXiv.org/abs/0903.1748}{\texttt{ arXiv:0903.1748}}.

\bibitem{Aad:2011qv}
\hrefCMSnoop {}{{ATLAS} Collaboration, ``Measurement of the production cross
  section for Z/$\Pgg^*$ in association with jets in $\Pp\Pp$ collisions at
  $\sqrt{s}$ = 7 TeV with the {ATLAS} detector'',} \textit{ Phys. Rev. D}
  \textbf{ 85} (2012) 032009,
  \href{http://dx.doi.org/10.1103/PhysRevD.85.032009}{\doi{10.1103/PhysRevD.85.032009}},
\href{http://www.arXiv.org/abs/1111.2690}{\texttt{ arXiv:1111.2690}}.

\bibitem{Chatrchyan:2011ne}
\hrefCMSnoop {}{{CMS} Collaboration, ``Jet production rates in association with
  W and Z bosons in $\Pp\Pp$ collisions at $\sqrt{s}$ = 7 TeV'',} \textit{
  JHEP} \textbf{ 01} (2012) 010,
  \href{http://dx.doi.org/10.1007/JHEP01(2012)010}{\doi{10.1007/JHEP01(2012)010}},
\href{http://www.arXiv.org/abs/1110.3226}{\texttt{ arXiv:1110.3226}}.

\bibitem{Aad:2013ysa}
\hrefCMSnoop {}{{ATLAS} Collaboration, ``Measurement of the production cross
  section of jets in association with a Z boson in $\Pp\Pp$ collisions at
  $\sqrt{s}$ = 7 TeV with the {ATLAS} detector'',} \textit{ JHEP} \textbf{ 07}
  (2013) 032,
  \href{http://dx.doi.org/10.1007/JHEP07(2013)032}{\doi{10.1007/JHEP07(2013)032}},
\href{http://www.arXiv.org/abs/1304.7098}{\texttt{ arXiv:1304.7098}}.

\bibitem{CMS-PAS-SMP-12-008}
\href {https://cds.cern.ch/record/1434360}{{CMS} Collaboration, ``Absolute
  Calibration of the Luminosity Measurement at CMS: Winter 2012 Update'',} CMS
  Physics Analysis Summary CMS-PAS-SMP-12-008, 2012.

\bibitem{Chatrchyan:2008zzk}
\hrefCMSnoop {}{{CMS} Collaboration, ``The {CMS} experiment at the {CERN}
  {LHC}'',} \textit{ JINST} \textbf{ 3} (2008) S08004,
\href{http://dx.doi.org/10.1088/1748-0221/3/08/S08004}{\doi{10.1088/1748-0221/3/08/S08004}}.

\bibitem{Chatrchyan:1369486}
\hrefCMSnoop {}{{CMS} Collaboration, ``Determination of jet energy calibration
  and transverse momentum resolution in {CMS}'',} \textit{ JINST} \textbf{ 6}
  (2011) P11002,
  \href{http://dx.doi.org/10.1088/1748-0221/6/11/P11002}{\doi{10.1088/1748-0221/6/11/P11002}},
  \href{http://www.arXiv.org/abs/1107.4277}{\texttt{ arXiv:1107.4277}}.

\bibitem{Pumplin:2002vw}
P.~Jonathan\hrefCMSnoop {}{ {et~al.}, ``New generation of parton distributions
  with uncertainties from global {QCD} analysis'',} \textit{ JHEP} \textbf{ 07}
  (2002) 012,
  \href{http://dx.doi.org/10.1088/1126-6708/2002/07/012}{\doi{10.1088/1126-6708/2002/07/012}},
\href{http://www.arXiv.org/abs/hep-ph/0201195}{\texttt{ arXiv:hep-ph/0201195}}.

\bibitem{Sjostrand:2006za}
\hrefCMSnoop {}{T.~Sj{\"o}strand, S.~Mrenna, and P.~Skands, ``{PYTHIA} 6.4
  physics and manual'',} \textit{ JHEP} \textbf{ 05} (2006) 026,
  \href{http://dx.doi.org/10.1088/1126-6708/2006/05/026}{\doi{10.1088/1126-6708/2006/05/026}},
\href{http://www.arXiv.org/abs/hep-ph/0603175}{\texttt{ arXiv:hep-ph/0603175}}.

\bibitem{Khachatryan:2010nk}
\hrefCMSnoop {}{{CMS} Collaboration, ``Charged particle multiplicities in
  $\Pp\Pp$ interactions at $\sqrt{s}$ = 0.9, 2.36, and 7 TeV'',} \textit{ JHEP}
  \textbf{ 01} (2011) 079,
  \href{http://dx.doi.org/10.1007/JHEP01(2011)079}{\doi{10.1007/JHEP01(2011)079}},
\href{http://www.arXiv.org/abs/1011.5531}{\texttt{ arXiv:1011.5531}}.

\bibitem{Alwall:2007fs}
J.~Alwall\hrefCMSnoop {}{ {et~al.}, ``Comparative study of various algorithms
  for the merging of parton showers and matrix elements in hadronic
  collisions'',} \textit{ Eur. Phys. J. C} \textbf{ 53} (2008) 473,
  \href{http://dx.doi.org/10.1140/epjc/s10052-007-0490-5}{\doi{10.1140/epjc/s10052-007-0490-5}},
\href{http://www.arXiv.org/abs/0706.2569}{\texttt{ arXiv:0706.2569}}.

\bibitem{Golonka:2003xt}
P.~Golonka\hrefCMSnoop {}{ {et~al.}, ``The tauola-photos-F environment for the
  {TAUOLA} and {PHOTOS} packages, release {II}'',} \textit{ Comput. Phys.
  Commun.} \textbf{ 174} (2006) 818,
  \href{http://dx.doi.org/10.1016/j.cpc.2005.12.018}{\doi{10.1016/j.cpc.2005.12.018}},
\href{http://www.arXiv.org/abs/hep-ph/0312240}{\texttt{ arXiv:hep-ph/0312240}}.

\bibitem{Re:2010bp}
\hrefCMSnoop {}{E.~Re, ``Single-top {$W$ $t$-channel} production matched with
  parton showers using the {POWHEG} method'',} \textit{ Eur. Phys. J. C}
  \textbf{ 71} (2011) 1547,
  \href{http://dx.doi.org/10.1140/epjc/s10052-011-1547-z}{\doi{10.1140/epjc/s10052-011-1547-z}},
\href{http://www.arXiv.org/abs/1009.2450}{\texttt{ arXiv:1009.2450}}.

\bibitem{Gleisberg:2008fv}
\hrefCMSnoop {}{T.~Gleisberg and S.~{H\"{o}che}, ``Comix, a new matrix element
  generator'',} \textit{ JHEP} \textbf{ 12} (2008) 039,
  \href{http://dx.doi.org/10.1088/1126-6708/2008/12/039}{\doi{10.1088/1126-6708/2008/12/039}},
\href{http://www.arXiv.org/abs/0808.3674}{\texttt{ arXiv:0808.3674}}.

\bibitem{Schumann:2007mg}
\hrefCMSnoop {}{S.~Schumann and F.~Krauss, ``A parton shower algorithm based on
  {Catani-Seymour} dipole factorisation'',} \textit{ JHEP} \textbf{ 03} (2008)
  038,
  \href{http://dx.doi.org/10.1088/1126-6708/2008/03/038}{\doi{10.1088/1126-6708/2008/03/038}},
\href{http://www.arXiv.org/abs/0709.1027}{\texttt{ arXiv:0709.1027}}.

\bibitem{Gleisberg:2008ta}
T.~Gleisberg\hrefCMSnoop {}{ {et~al.}, ``Event generation with {SHERPA} 1.1'',}
  \textit{ JHEP} \textbf{ 02} (2009) 007,
  \href{http://dx.doi.org/10.1088/1126-6708/2009/02/007}{\doi{10.1088/1126-6708/2009/02/007}},
\href{http://www.arXiv.org/abs/0811.4622}{\texttt{ arXiv:0811.4622}}.

\bibitem{Hoeche:2009rj}
\hrefCMSnoop {}{S.~{H\"{o}che}, F.~Krauss, S.~Schumann, and F.~Siegert, ``{QCD}
  matrix elements and truncated showers'',} \textit{ JHEP} \textbf{ 05} (2009)
  053,
  \href{http://dx.doi.org/10.1088/1126-6708/2009/05/053}{\doi{10.1088/1126-6708/2009/05/053}},
\href{http://www.arXiv.org/abs/0903.1219}{\texttt{ arXiv:0903.1219}}.

\bibitem{Nadolsky:2008zw}
P.~M. Nadolsky\hrefCMSnoop {}{ {et~al.}, ``{Implications of CTEQ global
  analysis for collider observables}'',} \textit{ Phys. Rev. D} \textbf{ 78}
  (2008) 013004,
  \href{http://dx.doi.org/10.1103/PhysRevD.78.013004}{\doi{10.1103/PhysRevD.78.013004}},
\href{http://www.arXiv.org/abs/0802.0007}{\texttt{ arXiv:0802.0007}}.

\bibitem{Melnikov:2006kv}
\hrefCMSnoop {}{K.~Melnikov and F.~Petriello, ``Electroweak gauge boson
  production at hadron colliders through $O(\alpha_s^2)$'',} \textit{ Phys.
  Rev. D} \textbf{ 74} (2006) 114017,
  \href{http://dx.doi.org/10.1103/PhysRevD.74.114017}{\doi{10.1103/PhysRevD.74.114017}},
\href{http://www.arXiv.org/abs/hep-ph/0609070}{\texttt{ arXiv:hep-ph/0609070}}.

\bibitem{Martin:2009iq}
\hrefCMSnoop {}{A.~D. Martin, W.~J. Stirling, R.~S. Thorne, and G.~Watt,
  ``Parton distributions for the {LHC}'',} \textit{ Eur. Phys. J. C} \textbf{
  63} (2009) 189,
  \href{http://dx.doi.org/10.1140/epjc/s10052-009-1072-5}{\doi{10.1140/epjc/s10052-009-1072-5}},
\href{http://www.arXiv.org/abs/0901.0002}{\texttt{ arXiv:0901.0002}}.

\bibitem{Czakon:2013goa}
\hrefCMSnoop {}{M.~Czakon, P.~Fiedler, and A.~Mitov, ``{Total} {Top-Quark}
  {Pair-Production} {Cross} {Section} at {Hadron} {Colliders} {Through}
  $\mathcal{O}(\alpha_S^4)$'',} \textit{ Phys. Rev. Lett.} \textbf{ 110} (2013)
  252004,
  \href{http://dx.doi.org/10.1103/PhysRevLett.110.252004}{\doi{10.1103/PhysRevLett.110.252004}},
\href{http://www.arXiv.org/abs/1303.6254}{\texttt{ arXiv:1303.6254}}.

\bibitem{Campbell:2011bn}
\hrefCMSnoop {}{J.~M. Campbell, R.~K. Ellis, and C.~Williams, ``Vector boson
  pair production at the {LHC}'',} \textit{ JHEP} \textbf{ 07} (2011) 018,
  \href{http://dx.doi.org/10.1007/JHEP07(2011)018}{\doi{10.1007/JHEP07(2011)018}},
\href{http://www.arXiv.org/abs/1105.0020}{\texttt{ arXiv:1105.0020}}.

\bibitem{Agostinelli:2002hh}
\hrefCMSnoop {}{S.~Agostinelli {et~al.}, ``{GEANT4}---a simulation toolkit'',}
  \textit{ Nucl. Instrum. Meth. A} \textbf{ 506} (2003) 250,
\href{http://dx.doi.org/10.1016/S0168-9002(03)01368-8}{\doi{10.1016/S0168-9002(03)01368-8}}.

\bibitem{Allison:2006ve}
\hrefCMSnoop {}{J.~Allison {et~al.}, ``{GEANT4} developments and
  applications'',} \textit{ IEEE Trans. Nucl. Sci.} \textbf{ 53} (2006) 270,
\href{http://dx.doi.org/10.1109/TNS.2006.869826}{\doi{10.1109/TNS.2006.869826}}.

\bibitem{CMS-PAS-PFT-09-001}
\href {https://cds.cern.ch/record/1194487}{{CMS} Collaboration, ``Particle-Flow
  Event Reconstruction in CMS and Performance for Jets, Taus, and \ETmiss'',}
  CMS Physics Analysis Summary CMS-PAS-PFT-09-001, 2009.

\bibitem{CMS-PAS-PFT-10-001}
\href {https://cds.cern.ch/record/1247373}{{CMS} Collaboration, ``Commissioning
  of the Particle-flow Event Reconstruction with the first {LHC} collisions
  recorded in the {CMS} detector'',} CMS Physics Analysis Summary
  CMS-PAS-PFT-10-001, 2010.

\bibitem{CMS-PAS-EGM-10-004}
\href {https://cdsweb.cern.ch/record/1279341}{{CMS} Collaboration, ``Electron
  reconstruction and identification at $\sqrt{s}=7$~TeV'',} CMS Physics
  Analysis Summary CMS-PAS-EGM-10-004, 2010.

\bibitem{Chatrchyan:2013dga}
\hrefCMSnoop {}{{CMS} Collaboration, ``Energy calibration and resolution of the
  {CMS} electromagnetic calorimeter in $\Pp\Pp$ collisions at
  $\sqrt{s}=7$\TeV'',} \textit{ JINST} \textbf{ 8} (2013) P09009,
  \href{http://dx.doi.org/10.1088/1748-0221/8/09/P09009}{\doi{10.1088/1748-0221/8/09/P09009}},
\href{http://www.arXiv.org/abs/1306.2016}{\texttt{ arXiv:1306.2016}}.

\bibitem{Cacciari:2007fd}
\hrefCMSnoop {}{M.~Cacciari and G.~P. Salam, ``Pileup subtraction using jet
  areas'',} \textit{ Phys. Lett. B} \textbf{ 659} (2008) 119,
  \href{http://dx.doi.org/10.1016/j.physletb.2007.09.077}{\doi{10.1016/j.physletb.2007.09.077}},
\href{http://www.arXiv.org/abs/0707.1378}{\texttt{ arXiv:0707.1378}}.

\bibitem{Cacciari:2008gp}
\hrefCMSnoop {}{M.~Cacciari, G.~P. Salam, and G.~Soyez, ``The anti-$k_t$ jet
  clustering algorithm'',} \textit{ JHEP} \textbf{ 04} (2008) 063,
  \href{http://dx.doi.org/10.1088/1126-6708/2008/04/063}{\doi{10.1088/1126-6708/2008/04/063}},
  \href{http://www.arXiv.org/abs/0802.1189}{\texttt{ arXiv:0802.1189}}.

\bibitem{Cacciari:2005hq}
\hrefCMSnoop {}{M.~Cacciari and G.~P. Salam, ``Dispelling the $N^{3}$ myth for
  the $k_t$ jet-finder'',} \textit{ Phys. Lett. B} \textbf{ 641} (2006) 57,
  \href{http://dx.doi.org/10.1016/j.physletb.2006.08.037}{\doi{10.1016/j.physletb.2006.08.037}},
\href{http://www.arXiv.org/abs/hep-ph/0512210}{\texttt{ arXiv:hep-ph/0512210}}.

\bibitem{Cacciari:2011ma}
\hrefCMSnoop {}{M.~Cacciari, G.~P. Salam, and G.~Soyez, ``{FastJet} user
  manual'',} \textit{ Eur. Phys. J. C} \textbf{ 72} (2012) 1896,
  \href{http://dx.doi.org/10.1140/epjc/s10052-012-1896-2}{\doi{10.1140/epjc/s10052-012-1896-2}},
\href{http://www.arXiv.org/abs/1111.6097}{\texttt{ arXiv:1111.6097}}.

\bibitem{CMS:2011aa}
\hrefCMSnoop {}{{CMS} Collaboration, ``Measurement of the inclusive W and Z
  production cross sections in $\Pp\Pp$ collisions at $\sqrt{s}=7$ TeV with the
  {CMS} experiment'',} \textit{ JHEP} \textbf{ 10} (2011) 132,
  \href{http://dx.doi.org/10.1007/JHEP10(2011)132}{\doi{10.1007/JHEP10(2011)132}},
\href{http://www.arXiv.org/abs/1107.4789}{\texttt{ arXiv:1107.4789}}.

\bibitem{arxiv:1105.1160}
\hrefCMSnoop {}{T.~Adye, ``Unfolding algorithms and tests using {RooUnfold}'',}
  (2011). \href{http://www.arXiv.org/abs/1105.1160}{\texttt{ arXiv:1105.1160}}.

\bibitem{nima:a372_469_481}
\hrefCMSnoop {}{A.~{H\"{o}cker} and V.~Kartvelishvili, ``{SVD} approach to data
  unfolding'',} \textit{ Nucl. Instrum. Meth. A} \textbf{ 372} (1996) 469,
  \href{http://dx.doi.org/10.1016/0168-9002(95)01478-0}{\doi{10.1016/0168-9002(95)01478-0}},
  \href{http://www.arXiv.org/abs/hep-ph/9509307}{\texttt{
  arXiv:hep-ph/9509307}}.

\bibitem{nima:a362_487_498}
\hrefCMSnoop {}{G.~D'Agostini, ``A multidimensional unfolding method based on
  Bayes' theorem'',} \textit{ Nucl. Instrum. Meth. A} \textbf{ 362} (1995) 487,
  \href{http://dx.doi.org/10.1016/0168-9002(95)00274-X}{\doi{10.1016/0168-9002(95)00274-X}}.

\bibitem{Chatrchyan:2012bra}
\hrefCMSnoop {}{{CMS} Collaboration, ``Measurement of the \ttbar production
  cross section in the dilepton channel in $\Pp\Pp$ collisions at $\sqrt{s}=7$
  TeV'',} \textit{ JHEP} \textbf{ 11} (2012) 067,
  \href{http://dx.doi.org/10.1007/JHEP11(2012)067}{\doi{10.1007/JHEP11(2012)067}},
\href{http://www.arXiv.org/abs/1208.2671}{\texttt{ arXiv:1208.2671}}.

\bibitem{Chatrchyan:2012ria}
\hrefCMSnoop {}{{CMS} Collaboration, ``Measurement of the \ttbar production
  cross section in $\Pp\Pp$ collisions at $\sqrt{s}=7$ TeV with lepton + jets
  final states'',} \textit{ Phys. Lett. B} \textbf{ 720} (2013) 83,
  \href{http://dx.doi.org/10.1016/j.physletb.2013.02.021}{\doi{10.1016/j.physletb.2013.02.021}},
\href{http://www.arXiv.org/abs/1212.6682}{\texttt{ arXiv:1212.6682}}.

\bibitem{Chatrchyan:2012bd}
\hrefCMSnoop {}{{CMS} Collaboration, ``Measurement of the sum of $\PW\PW$ and
  $\PW\cPZ$ production with \PW+dijet events in $\Pp\Pp$ collisions at
  $\sqrt{s}=7$ TeV'',} \textit{ Eur. Phys. J. C} \textbf{ 73} (2013) 2283,
  \href{http://dx.doi.org/10.1140/epjc/s10052-013-2283-3}{\doi{10.1140/epjc/s10052-013-2283-3}},
\href{http://www.arXiv.org/abs/1210.7544}{\texttt{ arXiv:1210.7544}}.

\bibitem{Chatrchyan:2012sga}
\hrefCMSnoop {}{{CMS} Collaboration, ``Measurement of the $\cPZ\cPZ$ production
  cross section and search for anomalous couplings in $2\ell2\ell^{\prime}$
  final states in $\Pp\Pp$ collisions at $\sqrt{s}=7$ TeV'',} \textit{ JHEP}
  \textbf{ 01} (2013) 063,
  \href{http://dx.doi.org/10.1007/JHEP01(2013)063}{\doi{10.1007/JHEP01(2013)063}},
\href{http://www.arXiv.org/abs/1211.4890}{\texttt{ arXiv:1211.4890}}.

\bibitem{Chatrchyan:2012saa}
\hrefCMSnoop {}{{CMS} Collaboration, ``Measurement of differential top-quark
  pair production cross sections in $\Pp\Pp$ collisions at $\sqrt{s}=7$ TeV'',}
  \textit{ Eur. Phys. J. C} \textbf{ 73} (2013) 2339,
  \href{http://dx.doi.org/10.1140/epjc/s10052-013-2339-4}{\doi{10.1140/epjc/s10052-013-2339-4}},
\href{http://www.arXiv.org/abs/1211.2220}{\texttt{ arXiv:1211.2220}}.

\bibitem{Buckley:2010ar}
A.~Buckley\hrefCMSnoop {}{ {et~al.}, ``Rivet user manual'',} \textit{ Comput.
  Phys. Commun.} \textbf{ 184} (2013) 2803,
  \href{http://dx.doi.org/10.1016/j.cpc.2013.05.021}{\doi{10.1016/j.cpc.2013.05.021}},
\href{http://www.arXiv.org/abs/1003.0694}{\texttt{ arXiv:1003.0694}}.

\bibitem{Alwall:2008qv}
\hrefCMSnoop {}{J.~Alwall, S.~de~Visscher, and F.~Maltoni, ``{QCD} radiation in
  the production of heavy colored particles at the {LHC}'',} \textit{ JHEP}
  \textbf{ 02} (2009) 017,
  \href{http://dx.doi.org/10.1088/1126-6708/2009/02/017}{\doi{10.1088/1126-6708/2009/02/017}},
\href{http://www.arXiv.org/abs/0810.5350}{\texttt{ arXiv:0810.5350}}.

\bibitem{Hoeche:2012yf}
\hrefCMSnoop {}{S.~{H\"{o}che}, F.~Krauss, M.~{Sch\"{o}nherr}, and F.~Siegert,
  ``{QCD} matrix elements + parton showers. {The NLO} case'',} \textit{ JHEP}
  \textbf{ 04} (2013) 027,
  \href{http://dx.doi.org/10.1007/JHEP04(2013)027}{\doi{10.1007/JHEP04(2013)027}},
\href{http://www.arXiv.org/abs/1207.5030}{\texttt{ arXiv:1207.5030}}.

\bibitem{CT10:2010}
H.-L. Lai\hrefCMSnoop {}{ {et~al.}, ``New parton distributions for collider
  physics'',} \textit{ Phys. Rev. D} \textbf{ 82} (2010) 074024,
  \href{http://dx.doi.org/10.1103/PhysRevD.82.074024}{\doi{10.1103/PhysRevD.82.074024}},
\href{http://www.arXiv.org/abs/1007.2241}{\texttt{ arXiv:1007.2241}}.

\bibitem{Berger:2010gf}
C.~F. Berger\hrefCMSnoop {}{ {et~al.}, ``{Vector Boson + Jets with BlackHat and
  SHERPA}'',} \textit{ Nucl. Phys. Proc. Suppl.} \textbf{ 205-206} (2010) 92,
  \href{http://dx.doi.org/10.1016/j.nuclphysbps.2010.08.025}{\doi{10.1016/j.nuclphysbps.2010.08.025}},
\href{http://www.arXiv.org/abs/1005.3728}{\texttt{ arXiv:1005.3728}}.

\bibitem{Alioli:2010qp}
\hrefCMSnoop {}{S.~Alioli, P.~Nason, C.~Oleari, and E.~Re, ``Vector boson plus
  one jet production in {POWHEG}'',} \textit{ JHEP} \textbf{ 01} (2011) 095,
  \href{http://dx.doi.org/10.1007/JHEP01(2011)095}{\doi{10.1007/JHEP01(2011)095}},
\href{http://www.arXiv.org/abs/1009.5594}{\texttt{ arXiv:1009.5594}}.

\bibitem{Ball:2011mu}
R.~D. Ball\hrefCMSnoop {}{ {et~al.}, ``Impact of heavy quark masses on parton
  distributions and {LHC} phenomenology'',} \textit{ Nucl. Phys. B} \textbf{
  849} (2011) 296,
  \href{http://dx.doi.org/10.1016/j.nuclphysb.2011.03.021}{\doi{10.1016/j.nuclphysb.2011.03.021}},
\href{http://www.arXiv.org/abs/1101.1300}{\texttt{ arXiv:1101.1300}}.

\bibitem{Berends:1989cf}
F.~A. Berends\hrefCMSnoop {}{ {et~al.}, ``Multijet production in \PW,\Z events
  at $\Pp\Pap$ colliders'',} \textit{ Phys. Lett. B} \textbf{ 224} (1989) 237,
\href{http://dx.doi.org/10.1016/0370-2693(89)91081-2}{\doi{10.1016/0370-2693(89)91081-2}}.

\end{thebibliography}\endgroup

\cleardoublepage \appendix\section{The CMS Collaboration \label{app:collab}}\begin{sloppypar}\hyphenpenalty=5000\widowpenalty=500\clubpenalty=5000\textbf{Yerevan Physics Institute,  Yerevan,  Armenia}\\*[0pt]
V.~Khachatryan, A.M.~Sirunyan, A.~Tumasyan
\vskip\cmsinstskip
\textbf{Institut f\"{u}r Hochenergiephysik der OeAW,  Wien,  Austria}\\*[0pt]
W.~Adam, T.~Bergauer, M.~Dragicevic, J.~Er\"{o}, C.~Fabjan\cmsAuthorMark{1}, M.~Friedl, R.~Fr\"{u}hwirth\cmsAuthorMark{1}, V.M.~Ghete, C.~Hartl, N.~H\"{o}rmann, J.~Hrubec, M.~Jeitler\cmsAuthorMark{1}, W.~Kiesenhofer, V.~Kn\"{u}nz, M.~Krammer\cmsAuthorMark{1}, I.~Kr\"{a}tschmer, D.~Liko, I.~Mikulec, D.~Rabady\cmsAuthorMark{2}, B.~Rahbaran, H.~Rohringer, R.~Sch\"{o}fbeck, J.~Strauss, A.~Taurok, W.~Treberer-Treberspurg, W.~Waltenberger, C.-E.~Wulz\cmsAuthorMark{1}
% ...
\vskip\cmsinstskip
% ...
1:~~Also at Vienna University of Technology, Vienna, Austria\\
2:~~Also at CERN, European Organization for Nuclear Research, Geneva, Switzerland\\
% ...

\end{sloppypar}
\end{document}